\documentclass[journal]{IEEEtran}

\usepackage[utf8]{inputenc}
\usepackage{multicol}
\usepackage{color}
\usepackage{lscape}
\usepackage{graphicx}
\usepackage{cite}
\usepackage{url}
\usepackage{etoolbox}
\usepackage{soul}
\graphicspath{ {./images/} }

\newtoggle{finalPaper}
\toggletrue{finalPaper}

\iftoggle{finalPaper} {
    \newcommand{\addtxt}[1]{#1}
    \newcommand{\change}[2]{#2}
    \newcommand{\rmvtxt}[1]{}
} {
    \newcommand{\addtxt}[1]{\textcolor{blue}{#1}}
    \newcommand{\change}[2]{\textcolor{blue}{\st{#1}}\textcolor{blue}{#2}}
    \newcommand{\rmvtxt}[1]{\textcolor{blue}{\st{#1}}}
}

\begin{document}

\title{Tackling Cyberattacks through AI-based Reactive Systems: A Holistic Review and Future Vision}

\author{\IEEEauthorblockN{Sergio Bernardez Molina$^{1,3}$, Félix Gómez Mármol$^1$}, Pantaleone Nespoli$^{1,2,*}$\\
\IEEEauthorblockA{$^1$Department of Information and Communications Engineering, University of Murcia, 30100, Murcia, Spain\\
\ \{sergio.bernardezm,pantaleone.nespoli,felixgm\}@um.es}\\
\IEEEauthorblockA{$^2$SAMOVAR, Télécom SudParis, Institut Polytechnique de Paris, 91120 Palaiseau, France}
\\
\IEEEauthorblockA{$^3$Indra Sistemas, Avenida de Bruselas 35, 28108 Alcobendas, Spain}
\thanks{\IEEEauthorrefmark{1}Corresponding author: Pantaleone Nespoli (email: pantaleone.nespoli@um.es)}
}

\maketitle

\begin{abstract}
There is no denying that the use of Information Technology (IT) is undergoing exponential growth in today's world. This digital transformation has also given rise to a multitude of security challenges, notably in the realm of cybercrime. In response to these growing threats, public and private sectors have prioritized the strengthening of IT security measures. In light of the growing security concern, Artificial Intelligence (AI) has gained prominence within the cybersecurity landscape. This paper presents a comprehensive survey of recent advancements in AI-driven threat response systems. To the best of our knowledge, the most recent survey covering the AI reaction domain was conducted in 2017. Since then, considerable literature has been published\addtxt{,} and therefore\addtxt{,} it is worth reviewing it. \change{By means of several shared features, each of the studies is compared on a common ground.Through an analysis of the research papers conducted on a standardized basis, this survey aims to unravel the complexities and opportunities of integrating AI into cyber defense. The conclusions drawn from this collective analysis provide a comprehensive snapshot of the evolving landscape at the intersection of AI and cybersecurity. This landscape underscores the growing significance of not only anticipating and detecting threats but also responding to them effectively. Additionally, from these reviews, various research challenges for the future are presented. These challenges serve as a roadmap for researchers and practitioners in the field of AI-integrated reactive strategies.}{In this comprehensive survey of the state of the art reaction systems, five key features with multiple values have been identified, facilitating a homogeneous comparison between the different works. In addition, through a meticulous methodology of article collection, the 22 most relevant publications in the field have been selected. Then each of these publications has been subjected to a detailed analysis using the features identified, which has allowed for the generation of a comprehensive overview revealing significant relationships between the papers. These relationships are further elaborated in the paper, along with the identification of potential gaps in the literature, which may guide future contributions. A total of seven research challenges have been identified, pointing out these potential gaps and suggesting possible areas of development through concrete proposals.}  

\end{abstract}\begin{IEEEkeywords}
Artificial Intelligence, Cybersecurity, Reaction, Countermeasure, Survey 
\end{IEEEkeywords}

\section{Introduction}

The use of Information Technology (IT) is growing exponentially on an annual basis. An increasing number of services are adopting IT solutions to automate and enhance various processes. As a result, human beings are becoming increasingly dependent on technology. In this sense, the use of IT can add many facilities and advantages to society~\cite{advantages} in terms of services, both essential and merely recreational. However, this increase in the use of IT also poses many challenges, not only in terms of scalability and technical challenges but also in terms of security, since cybercrime~\cite{measuring} is also on the rise year after year. Specifically, cybercrime is a form of crime committed within the digital world. It includes a wide range of illegal practices, such as data theft~\cite{dataTheft}, hacking~\cite{hacking}, online harassment~\cite{harassment}, cyber espionage~\cite{espionage}, and distribution of malware and viruses~\cite{distribution}, among others. 

Thus, various recent reports~\cite{report1,report2} reflect cybercrime as one of the most lucrative illegal markets worldwide as of today. As a result, interest has expanded, triggering a trend towards individuals using the online environment for malicious purposes. These malefactors prioritize personal gain, often at the expense of IT users. Consequently, private organizations play a pivotal role in orchestrating these malicious activities. Some of them establish independent entities known as Advanced Persistent Threats (APTs)~\cite{APT}, while others may even carry out state-sponsored actions against foreign nations~\cite{stateSponsored}. As a matter of fact, recent activities provide a clear reflection of this phenomenon, especially in the recent war between Russia and Ukraine~\cite{russiaUCranian}. The computing or technological world has gained immense importance, evident in its influence on the military and critical infrastructure sectors~\cite{military,criticalSurvey}. Notably, in these domains, cyber-attacks have the potential to inflict greater impact than certain physical assaults. Such a phenomenon indicates a gradual convergence between the cyber world and the physical world, with cyber-attacks gaining ever-growing significance~\cite{cyberPhysical}. They can be compared, in terms of severity and impact, to actual physical attacks. 

Such malicious actions are reflected in frequent cyber-attacks affecting every connected asset. For example, Denial-of-Service (DoS)~\cite{DOS} attack can be coordinated by criminal organizations, states, or groups of self-interested individuals, resulting in a Distributed Denial-of-Service (DDoS) attack~\cite{DDOS}, and cause extensive damage to entire populations by attacking critical infrastructures such as the energy, water supply or transport sectors. This is why companies and states are investing more and more budget~\cite{budget} every year to increase the IT security of their systems and the awareness of employees susceptible to attacks. Nevertheless, studies such as~\cite{healthcare} show that these budgets are not enough, leaving systems exposed and unprotected. 

Given the alarming rise in cyber threats, the need to fortify security measures has become a top priority for both the public and private sectors. Consequently, the technology industry has been forced to invest heavily in research and development focused on enhancing systems designed to detect, prevent, and react to various types of cyber-attacks. A prominent example of these security tools is Intrusion Detection Systems (IDS)~\cite{IDS}, playing a vital role in monitoring systems and networks to identify and promptly report potential threats to security operators. These systems are often complemented by event management tools, such as Security Information and Event Management (SIEM)~\cite{SIEM,SiemLeo} systems, providing real-time insights into potential security risks. Furthermore, the integration of IDS with Intrusion Prevention System (IPS)~\cite{justIPS} and Intrusion Response System (IRS)~\cite{justIRS} has emerged as a comprehensive approach to strengthen defenses against the ever-evolving landscape of cyber threats.

The necessity to strengthen security measures is heightened by the emergence of Artificial Intelligence (AI)~\cite{emergingAI}, one of the cutting-edge advancements employed in the field of cybersecurity. AI is a leading technology that aims to create an automated intelligence capable of responding like human intelligence~\cite{aidefinition}. To achieve this objective, machines are trained using learning algorithms to acquire precise knowledge and develop cognitive capabilities. AI methods heavily rely on sophisticated algorithms, empowering machines to process vast amounts of data, recognize patterns~\cite{patterns}, make informed decisions~\cite{informedDecision}, and respond intelligently to various tasks and challenges. By emulating human intelligence, AI technology holds the potential to revolutionize industries, from healthcare and finance to cybersecurity and beyond. This technique has come a long way in recent years, powered by advances in Machine Learning (ML), natural language processing, Neural Network (NN)~\cite{neuralNetworks}, and other related fields~\cite{progressAI}. The potential use of AI in cybersecurity has led to a notable trend focused on integrating systems such as IRS using AI capabilities~\cite{id6}. The aim is to develop tools that autonomously select and apply the best countermeasures against cyber-attacks with efficiency. Leveraging the power of AI, these tools can effectively identify emerging threats, propose appropriate countermeasures and significantly reduce the impact of these threats on organizations and systems~\cite{id22}. 

The conjunction between AI and cybersecurity reaction systems can bring a powerful combination that amplifies the effectiveness of \change{cyber defense}{cybersecurity}  mechanisms. For example, where IRS focuses on real-time threat monitoring, analysis, and incident response, AI increases these capabilities by taking advantage of advanced analytics, ML, and predictive models. In this sense, the collaboration enables the identification of sophisticated attack patterns, the prediction of unknown future threats, as well as the generation of actionable perspectives to inform decision-making processes~\cite{decisionMaking}. 

Some researchers have recognized the significance of incorporating AI techniques into \change{cyber defense}{cybersecurity} mechanisms~\cite{Mohanty2018}. Furthermore, numerous papers have proposed a variety of approaches to develop more effective countermeasure selection mechanisms that use AI. Therefore, it is crucial to review and extend the findings obtained from the studies by integrating the latest advances in AI and capturing the knowledge gained from subsequent research efforts. This is why it is important to review the state of the art of threat response systems using AI so that all developments or approaches in the field can be brought into a unified vision.  

Acknowledging the abundance of perspectives on countermeasures, reaction strategies, and selection mechanisms, this paper aims to unify these various perspectives under a single, comprehensive picture. By doing so, the main objective is to create a survey that unifies the intricate threads of AI-based cybersecurity. In the end, the goal is to provide a cohesive overview under which diverse methodologies can converge, leading to a more robust and consolidated approach to cybersecurity. Besides, it is noteworthy that, throughout history, the emphasis in cybersecurity has been predominantly on threat detection, often overlooking the importance of reactive response strategies~\cite{IDSvsIPS,PAPADAKI200415}. This imbalance could be attributed to the appeal of anticipation and prevention, which have gained more attention due to their potential to stop threats before becoming reality. Still, the consequences of neglecting the reaction aspect are significant. By clarifying this imbalance, it sheds light on the importance of promoting a comprehensive overview that not only anticipates and detects threats, but also responds to them fast and effectively.

There have been previous studies~\cite{Survey1,INAYAT201653,Survey2,Survey3,BASHENDY2023102984,electronics11071072} that unify previously conducted work investigating the automaticity of countermeasure selection and reaction strategies. However, due to the rapid pace of advancements in the field of reaction mechanisms using AI, there is a need for an up-to-date study to consolidate existing knowledge, establish benchmarks, and compare different modern approaches. As far as we are aware, the latest survey addressing the field of AI reaction was carried out in 2017. Therefore, by conducting a comprehensive study, a holistic understanding of the current state of the art can be gained, research gaps can be identified, and valuable information can be provided for researchers, practitioners, and organizations working to improve cybersecurity defenses through AI-based reaction systems. \addtxt{In addition to the above, it is important to emphasize that this paper extends beyond a narrow focus on threat response. Instead, it aims to explore the convergence of cybersecurity and AI, reflecting on their current state of integration. Consequently, our analysis encompasses not only IRS but also evaluates works featuring AI-driven reactive mechanisms such as algorithms, frameworks, and architectures. By addressing both challenges and advancements in this domain, our study illuminates the complexities and opportunities inherent in safeguarding systems from cyber threats.}

\begin{figure}[h!]
    \centering
    \includegraphics[width=\columnwidth]{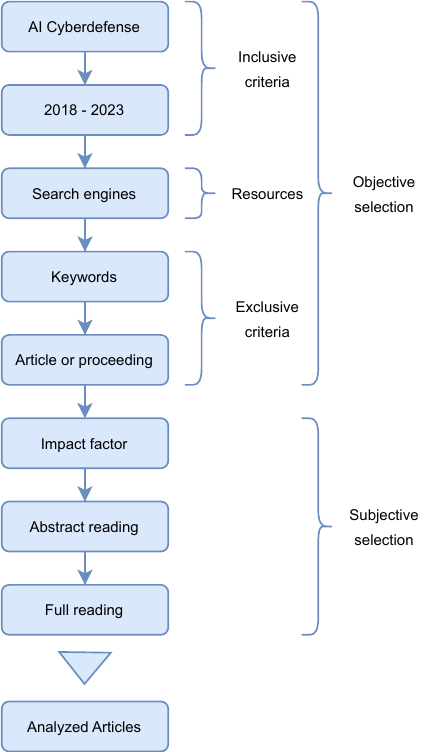}
    \caption{\addtxt{Methodology used for the selection of the papers to be analyzed in search of a comprehensive overview of the state of the art.}}
    \label{fig:methodologySelection}
\end{figure}

\change{For this reason, the objective of this paper is to conduct a comprehensive survey and analysis of AI-based reaction systems. To do so, the most remarkable 22 papers (to the best of our knowledge) from the last 5 years have been analyzed, comparing them in an equivalent basis across 5 selected features that have been considered relevant to the world of reaction. }{In order to carry out this study, a recent state-of-the-art recollection work has been carried out, following the methodology described in \figurename~\ref{fig:methodologySelection}. The diagram illustrates two types of selection: objective and subjective. In particular, the research started with a literature search based on inclusive criteria, focusing on the intersection of cyber defense and AI between 2018 and 2023. This criterion was applied in reputable search engines such as Google Scholar, Science Direct, IEEE Xplore, and Research Gate, with additional keywords such as ``reaction'', ``countermeasure'', ``cyber defense'', ``AI'', and ``autonomous countermeasure system''. During the initial selection phase, materials not directly published in journals or conferences were excluded. Preference was given to publications with a higher impact factor to ensure the inclusion of high-quality academic work.}

\addtxt{After objective filtering, the abstracts of the pre-selected papers were reviewed to identify those most aligned with the research objectives. This subjective assessment took into account factors such as relevance, novelty, and methodological rigor. Subsequently, each selected article was subjected to an exhaustive reading that allowed us to identify the most outstanding articles of the last five years. In total, 22 exemplary articles were selected.}

\change{Moreover, the paper aims to individually examine and impartially compare these works. In doing so, it is intended to highlight emerging challenges and outline possible research directions for future research in the field of AI-integrated reactive strategies.  Furthermore, the paper aims to locate relevant related studies that have the potential to improve the effectiveness of such systems in future efforts.}{Our research revealed a critical gap in current field studies, i.e., the limited exploration of AI-based solutions for reactive countermeasures. Our analysis of the state of the art revealed significant underdevelopment in the field of reactive defense strategies, leaving a heavy reliance on human intervention amongst the increasing complexity of modern cyber threats. Thus, the advent of AI offers an important opportunity that is crucial to study and expand.}

\addtxt{Consequently, this study was conducted to fill such a gap and shed light on the current state of such systems to ultimately identify their main challenges, trying to unify the research found so far under the same common characteristics and to create a comprehensive picture. By analyzing existing work and outlining future challenges, the paper aims to motivate the research community to further explore and develop AI-integrated reactive strategies in cybersecurity. Therefore, the paper aims to contribute significantly to the field by providing the following: }
\addtxt{
\begin{enumerate}
    \item Related studies are identified and synthesized to enrich community knowledge, providing guidance based on the strengths and weaknesses of the selected papers.
    \item Side-by-side comparisons to elucidate key trends and emerging challenges within the domain of AI-integrated cybersecurity strategies.
    \item Outlining potential research directions to stimulate further inquiry and innovation in this critical area of study.
\end{enumerate}
}

The rest of the paper has been organized in the following manner. In Section~\ref{Section:Background} certain concepts are introduced as foundational information for the benefit of the reader. It presents the key concepts essential to understand the research in this area. Section~\ref{Features} introduces a list of features used in the research. These features are later used for a comparative analysis between papers. Section~\ref{Section:Survey} provides each of the papers analyzed for this survey in sub-sections. An overview of each proposal is given and their practicality is discussed based on the features. Section~\ref{ComparativeExploration} a comparison detailing the overview from the papers analyzed is detailed. Next, Section~\ref{Section:ResearchChallenge} lists the potential directions for future research identified through the comparative analysis of the previous sections. Finally, Section~\ref{Section:Conclusion} concludes the paper by presenting its conclusions.

\section{Background}\label{Section:Background}

For the sake of the reader, the main objective of this section is to cover several key concepts that form the basis of AI-powered threat response systems. Particularly, understanding these concepts is essential to understanding the context and importance of this research ecosystem. As discussed previously, the increase in the number of cyber-attacks is tangible. This is why systems such as IDS, which are dedicated to detecting threats, are used. And, it is because of the inability of IDS to react that they are combined with response systems such as IPS or IRS. 

Previously, cyber-attack detection and prevention systems heavily relied on static techniques such as rule-based analysis~\cite{rulebased1} or approaches built upon hash-based analysis~\cite{static1}  or the normal behavior of applications~\cite{static2}. However, with the advancement of technology and the increasing sophistication of cyber threats, static detection has become inefficient~\cite{static3}. Therefore, there is a growing need for more advanced and dynamic approaches to effectively safeguard against evolving attack vectors, that can cover modern attacks. 

On the one hand, IPS is a network security tool that aims to prevent cyber threats by analyzing network traffic and taking a proactive and preventive approach to security. IPSs try to prevent attacks by identifying and blocking suspicious activities before they can cause damage proactively.  There are different approaches to the design of these systems. The approach in~\cite{ipsexample1} proposes a self-organizing incremental NN with a support vector machine to identify threats in the systems. The novelty of the solution relies on the structure that does not rely on signatures or rules for the mitigation of known attacks. Another example is~\cite{ipsexample2} which proposes an OpenFlow-based IPS for cloud environments. The solution depends on the use of Attack graph (AG)~\cite{attackgraph}  and an attack analyzer in the cloud. It is worth remarking that an AG is a representation of all possible paths of attack against a cybersecurity network.

On the other hand, IRS is another critical component of cybersecurity that should be part of any overall security plan. IRS is designed to identify and respond to security incidents in real time, thereby reducing the impact of the attack significantly. An effective IRS not only identifies potential threats but also executes an appropriate reaction, such as isolating affected systems or blocking malicious traffic. By responding fast to emerging threats, the IRS may significantly reduce the risk of data loss, network downtime, and reputational damage. 

\begin{figure}[h!]
    \centering
    \includegraphics[width=\columnwidth]{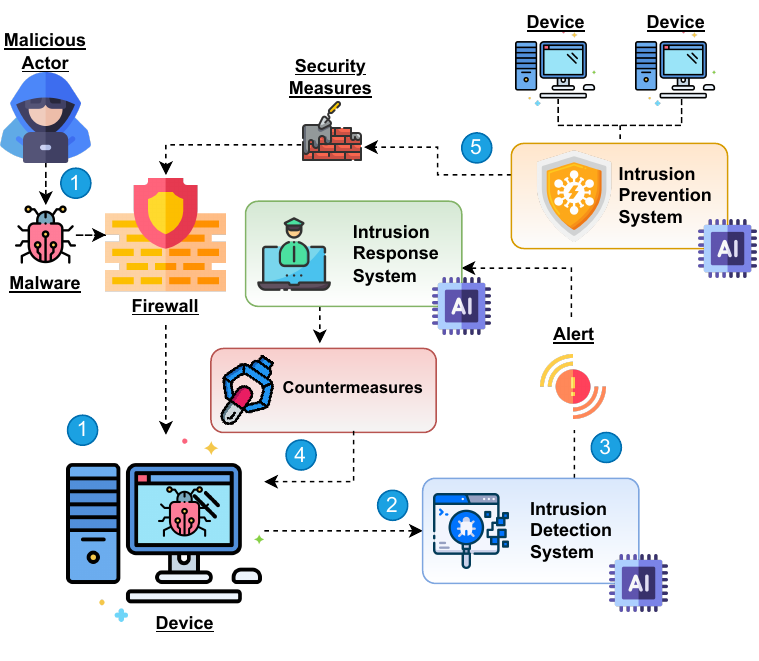}
    \caption{Prevention, response\addtxt{,} and detection AI-based systems scenario. }
    \label{fig:puzzle}
\end{figure}

\rmvtxt{For a better comprehension understanding of each of these systems (IDS, IPS and IRS), Fig. 2 has been designed. In this figure, the different components are graphically displayed. The scene is represented in some steps, that correspond to the numbers associated in the figure: 1) The scenario describes an attack from a malicious actor where a device is infected. 2) After infection, the IDS would be in charge of detecting the malware on the infected device. 3) Once identified the malware, it would send an alarm to the IRS. In this case, the IRS would be in charge of the incident response. 4) After analyzing the alert it receives from the IPS, the IRS would then select the most effective countermeasure and implement it. 5) On the other hand, an IPS is constantly analyzing the network and proactively generating security measures. }
\addtxt{For a better comprehension understanding of an AI-driven response system, in conjunction with other essential components of a network (IDS, IPS), \figurename~\ref{fig:puzzle} has been designed. In this figure, the different components are graphically displayed. The logical flow is divided in steps, that correspond to the numbers associated with the figure: }

\addtxt{
\begin{enumerate}
    \item The scenario describes an attack from a malicious actor where a device is infected.
    \item After infection, the IDS would be in charge of detecting the malware on the infected device. AI is particularly useful at this stage, as it can analyze vast amounts of data in real time, recognizing patterns and anomalies that might indicate an intrusion.
    \item Once identified the malware, the IDS would send an alarm to the IRS. The IRS then analyzes the nature of the threat and selects the most suitable and effective countermeasure. AI's role here is crucial, as it can analyze historical data and current threats to suggest optimal responses.
    \item After analyzing the alert it receives from the IPS, the IRS would then select the most effective countermeasure and implement it.
    \item Meanwhile, an IPS is constantly analyzing the network and proactively generating security measures. The AI-powered IPS acts as a proactive barrier, preventing potential intrusions before they can exploit vulnerabilities in the network.
\end{enumerate}
}

Technologies such as AI can offer a new approach that leverages the latest advancements to create highly efficient systems\cite{AIcyberdefense,useofAI}. These new developments have the potential to enhance on many levels, for instance in the fields of medicine~\cite{AIMedicine}, robotics~\cite{IARobotics}, finance~\cite{AIFinance}, and transport~\cite{AITransport}, among others.

\rmvtxt{When it comes to cyber defense, the deployment of these advances can help for example in improving the detection, prevention, and response systems discussed above.}

For instance, one main advantage of AI in cybersecurity is its ability to analyze large volumes of data and identify patterns and anomalies that may be indicative of a cyber-attack~\cite{aicapabilitiesdetect}. For example, algorithms using ML can be trained on large amounts of datasets to recognize different patterns of known attacks and detect previously unseen threats, such as ``zero-days" vulnerabilities~\cite{0dayai}. As already said, traditional approaches to security rely on static, rule-based systems that are unable to handle the increasing complexity and sophistication of cyber threats. AI, on the other hand, has a strong ability to learn and adapt to new threats, making it a powerful cybersecurity tool.

\addtxt{When it comes to cybersecurity, the deployment of these advances can help for example in improving the detection, prevention, and response systems discussed above.}

\addtxt{Particularly, AI can potentially enhance IDS by enabling it to analyze extensive datasets, thereby improving detection accuracy and minimizing false positives through behavioral analysis. By analyzing the behavior of users and systems, AI algorithms could learn normal behavior patterns, allowing them to identify deviations or anomalies that suggest possible intrusions or malicious activities, even in encrypted traffic \cite{Bakhshi2021}. Additionally, an IDS could be integrated with threat intelligence feeds, continually updating its understanding of emerging threats and attack vectors.}
 
\addtxt{Similarly, IPS systems can benefit from AI by enabling dynamic adjustment of security policies or cybersecurity strategies, informed by evolving threat landscapes and historical data insights. With AI-driven IPS, rapid discrimination between benign network activities and malicious intrusions becomes possible. Real-time analysis of network traffic by AI algorithms enables the dynamic adaptation of firewall rules and access controls to counter emerging threats. Similarly, as in IDS systems, AI can support IPS systems by identifying zero-day threats through analysis of behavioral patterns and anomalies.Integrating AI into IPS systems can also present a significant advantage in handling massive amounts of data. AI algorithms excel at processing and analyzing large data sets, allowing IPS systems to extract valuable information from the growing volume of network traffic and security logs.}

\addtxt{Within the IRS ecosystem, by deploying natural language processing techniques, AI could facilitate the rapid understanding and categorization of incident reports, accelerating the prioritization and allocation of resources for response efforts. AI-driven automation accelerates the execution of pre-defined response procedures, enabling security incidents to be quickly contained and mitigated. In addition, AI-powered IRS could automate the initial triage of security incidents by analyzing incoming alerts and prioritizing them based on severity, impact, and likelihood of exploitation. In addition, AI in an IRS system could perform predictive analytics to anticipate future security incidents by leveraging historical data and current threat intelligence. Another benefit of this integration could be the ability to quickly and efficiently select the optimal set of countermeasures to address threats in real time.}

In addition, the autonomous nature of these AI-based threat response systems represents a significant advantage. They are capable of continuously monitoring network activity, proactively identifying potential vulnerabilities, and promptly responding to mitigate risks~\cite{AInetwork}. By automating threat detection and response processes, organizations can greatly enhance their security posture and reduce the time gap between threat identification and resolution. 

Once the benefits of AI have been explored, it is also important to recognize that this technology comes with certain limitations and risks. The study conducted in~\cite{disadvantageIA} discusses some of these disadvantages. Among these challenges, one involves the inherent unpredictability of AI, which concerns the possibility that AI responses or actions may deviate from the intended goals or objectives. Furthermore, authors highlight that the creation of a super-intelligent AI could result in more harm than benefit when exploited by malicious entities. This underlines the urgency of managing the unpredictability that is inherent in AI systems. Another study~\cite{effectAICyber} identifies additional negative applications that can be obtained by AI exposing potential risks and ethical concerns. For example, AI can be used as a tool to generate deep fakes~\cite{deepfakeia}, which can serve as a powerful disinformation weapon used by companies or states, capable of impacting vast numbers of individuals due to the widespread reach of the Internet. Another use can be for password-guessing tools~\cite{passwordCrackIA} or the impersonation of humans on social media platforms~\cite{impersonatingIA}. Through sophisticated algorithms, AI can replicate the behavior of real individuals, seeking personal advantages or even financial gain. The last of the malicious tools proposed in the study is the use of AI for Supported Hacks, where AI is trained to execute various attacks, in this case, Wi-Fi attacks, through the Pwnagotchi 1.0.0 tool~\cite{pwnagotchi}, to deauthenticate users from wireless networks. 

When integrating AI into threat response systems\addtxt{, such as an autonomous IRS}, there are various goals to accomplish. First, is to achieve an autonomous selection of measures to counter ongoing cyber threats. A second goal is the ability that involves the restoration of the system to a stable state. Therefore, these systems must have predefined guidelines to prevent and restore from diverse identified attacks. The preventive defense measures are called countermeasures. Although it is important to note that there can also be reactive countermeasures and not only preventive countermeasures. Countermeasures can be defined as the actions that must be taken to effectively stop an attempted attack. In the field of computing and telecommunications, examples of countermeasures are ``shutting down specific ports", ``cutting off external communications", ``deleting files" or ``terminating running processes", etc.  

Numerous studies are dedicated to identify optimal countermeasures for different environments, such as remote vehicle communication~\cite{threatsCountermeasures}, smart grid~\cite{smartGrid,khoei2022comprehensive}, or robotics~\cite{roboticsSurvey}. As a result, once the countermeasures that can be optimal to mitigate the risk of a detected threat have been defined, it is important to establish whether the response system should act autonomously or rely on the prior decision of a human operator in the loop. This is especially crucial in sensitive environments, such as military or sensitive information environments, where the measures taken may have a significant impact. For example, in a military environment, actions are carefully measured and should not be spontaneous~\cite{militaryDirections}, as the goal is to maintain a high degree of control in any given situation. This is where AI faces one of its biggest challenges to overcome, which are unpredictability and lack of explainability. Therefore, one can conclude that the degree of autonomy of a response system will depend on the specific context in which it is deployed.

\section{Features}\label{Features}
 
As already seen, there is existing research concerning the integration of AI alongside in reaction systems. In an effort to identify any gaps or emerging patterns in the architectural framework for context-specific solutions, an attempt has been made to distinguish specific common features that can be used for a comparative analysis of research in this area. Such an effort is intended to facilitate the identification of promising methodologies, enabling their consolidation for future research or, at the very least, to provide a baseline foundation for the initiation of a more unified approach.

\begin{figure}
    \centering
    \includegraphics[]{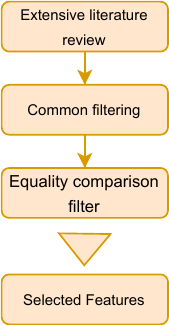}
    \caption{\addtxt{The methodology employed for selecting the comparative attributes to analyze the selected papers.}}
    \label{fig:FeaturesWorkflow}
\end{figure}

\addtxt{The selection of features to compare the surveyed papers was a meticulous process based on a structured methodology shown in \figurename~\ref{fig:FeaturesWorkflow}. In particular, we began by conducting an extensive literature review to understand the research landscape relevant to the reaction ecosystem. This review revealed many characteristics that were found in a variety of studies. These features were then subjected to a filtering process aimed towards isolating those consistently present in the papers studied. This initial filtering ensured that only the most prevalent and relevant features were considered. In addition, a further filter was applied to select those characteristics that would facilitate a meaningful comparison in the context of this study.}

It is essential to remark that these chosen features constitute one potential proposal among various alternatives. Other features could potentially be gathered and analyzed, but we firmly consider that the selected are the most advantageous choices. This conviction is based on the fact that these features represent in a simple and graphic way the instruments or methodologies used in the proposal of each of the selected papers. With these features, a reader can gain an overview of the most common approaches used in the world of reaction systems, which scenarios are most likely to be addressed, and even the rationale behind the algorithms for the search for countermeasures. 

In the following sections, each of the features is described, explaining each of the values that have been considered most relevant in our perspective. \addtxt{Furthermore, we elaborate on the impact that these features could have on responses to cyber-attacks and AI-based threat response systems. The aim is to establish the connections and effects between the identified characteristics and their importance for effectively mitigating cyber threats. Through this approach, a comprehensive understanding of how each characteristic contributes to the overall cyber defense strategy and resilience against evolving threats is provided.} The rest of the values are listed in section~\ref{featuresSummary} and are further explained in  section~\ref{Section:Survey}. 

\subsection{Methodology}

The ``Methodology'' feature attempts to represent the approach or procedure used in each of the investigations. It illustrates the techniques that the researchers have applied to solve a defined challenge. This challenge can be the selection of countermeasures, the generation of security policies, the identification of threats, etc. In particular, this feature has been selected as it helps to recognize the most commonly used techniques, and may even serve to identify untested methodologies that can be explored.

Methodology is an important attribute, as it defines the strategy followed by the authors of the analyzed works. Specifically, it involves the representation of the main technique that has been followed. The papers selected in the survey present a range of different methodologies. Some of the most significant examples are shown below.

\begin{itemize}
    \item \textbf{\change{ML}{Machine Learning (ML)}}: An evolving discipline in which computer algorithms are designed to emulate human intelligence through computer-based learning based on a digital environment~\cite{machineLearning}. From this digitalized context, an ML approach could be used to identify and learn from patterns within the collected data. \change{In particular,}{In particular, in order to counteract cyber-attacks,} it could be used in a reactive framework\addtxt{,} for example\addtxt{,} to dynamically adjust countermeasures according to the latest attack trends and tactics, or optimize resource allocation by calculating the costs of different countermeasures. \addtxt{In addition, ML algorithms can play a crucial role in optimizing the allocation of resources by calculating the costs associated with different countermeasures. }
    \item \textbf{Bayesian Networks}: These are specific types of graphical models. The specific type are Directed Acyclic Graphs (DAG)~\cite{directedAcyclic, Stephenson:82584}. In other words, all the edges of the graph are directed and there are no cycles. Thus, to solve the tasks, probabilistic relations are used as a working model on these graphs. Specifically, it could be used for risk calculations against the selection of countermeasures in a reactive environment. \addtxt{For instance, it could analyze incoming data from security sensors, assessing anomalies and patterns indicative of cyber threats. In addition, they can facilitate the development of effective incident response plans by modeling the dependencies between different response actions and their outcomes.} \figurename~\ref{fig:BayesianNetworkImage} \change{shows an example of a Bayesian network. It shows the probabilities for each branch and the different nodes of the network representing different states. }{ illustrates a Bayesian network designed to model various states and probabilities associated with the failure of a system, either due to malware exploitability or hardware failure. The network consists of several nodes representing different states within a security incident. Each branch of the network signifies a potential outcome, while the nodes encapsulate the probability of these outcomes. Through this representation, one can discern the potential scenarios that may unfold during a security breach, along with their associated probabilities.}

\begin{figure}
    \centering
    \includegraphics[width=\columnwidth]{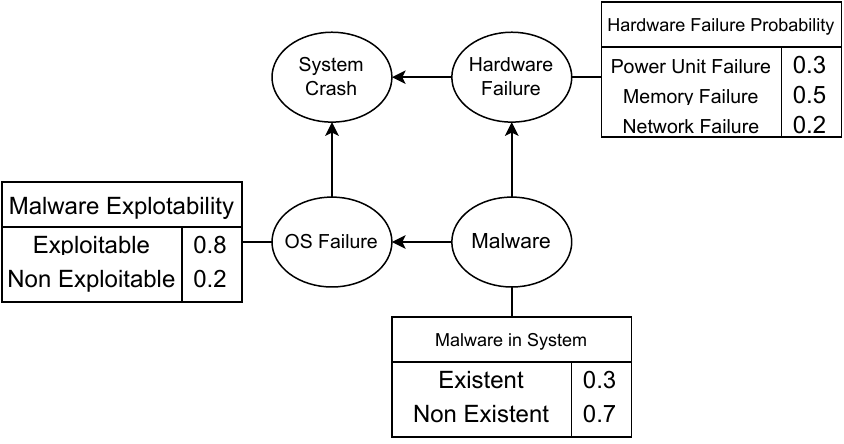}
    \caption{\change{Graphical Bayesian Network example.}{Probabilistic analysis of computer system failures represented through a graphical Bayesian Network.}}
    \label{fig:BayesianNetworkImage}
\end{figure}

    \item \textbf{Attack Graphs}: Graphical description used in cybersecurity to represent and visualize potential attack paths and sequences that an adversary could follow to compromise a computer system or network~\cite{attackgraph}. \change{It can be used to analyze vulnerabilities and countermeasures.}{Attack graphs are a versatile tool, capable of analyzing vulnerabilities and countermeasures. Thereby, these graphs can help to identify interconnected vulnerabilities and their impact on the overall security posture. Consequently, organizations can develop accurate and tailored mitigation strategies to effectively address these weaknesses.} \figurename~\ref{fig:ExampleAttackGraph} graphically depicts an example attack represented through an attack graph. \addtxt{In the graph, an example of how the attack could start via a browser can be seen. Through spear phishing a user account, and through the execution of the attack, the computer could be infected, reaching the operating system itself.} 

\begin{figure}
    \centering
    \includegraphics[width=\columnwidth]{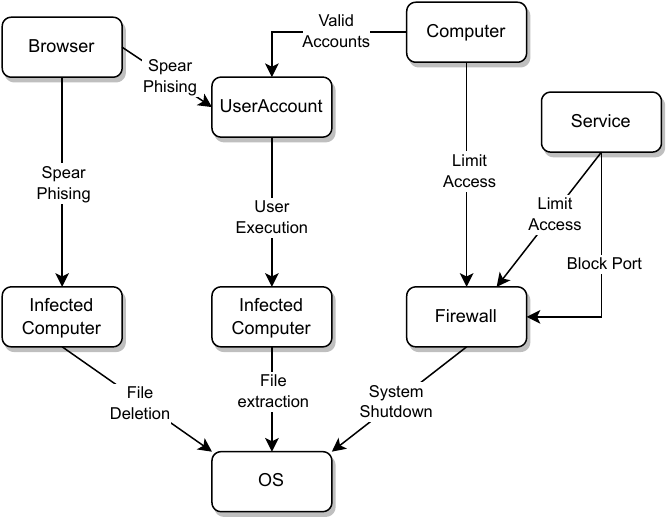}
    \caption{\change{Attack Graph example}{Graphical representation of an attack against an operating system using an Attack Graph.}}
    \label{fig:ExampleAttackGraph}
\end{figure}
    
\end{itemize}

\subsection{Algorithm}

The ``Algorithm" feature provides a more granular view of the technical tools and procedures used in the research. It helps to identify the specific algorithms commonly employed by researchers as well as potential, unexplored alternatives. This feature narrows down the focus to the computational and algorithmic components of the research, offering a more detailed examination of the technical aspects of problem-solving.

It is essential to note that while the "Algorithm" section goes into the specific computational techniques, it is distinct from the "Methodology" section. Unlike "Algorithm", which focuses mainly on computational and algorithmic elements, "Methodology" provides a broader representation of the general approach applied in the articles. 

The range of values in this feature is wider, of which some of the relevant examples that have been identified are:  

\begin{itemize}
    \item \textbf{Neural Networks (NN)}: Self-taught learning algorithm inspired by the functioning of the biological nervous system, particularly neurons in the human brain~\cite{neuralNetworks2}. It consists of interconnected nodes that interact and process information to be able to recognize patterns, classification, and regression, among others. \figurename~\ref{fig:NeuralNetworkImage}  displays an example of NN, in which three layers are detailed. The first layer is in charge of receiving the data, in the hidden layer particular processes are carried out, which can finally be seen in the output layer that delivers the desired result. \change{This type of network can be used in reactive systems to adapt to changing threats by the countermeasure selection system. Furthermore, it could be used for threat classification to select the most optimal countermeasures to be applied.  }{This type of network is useful in reactive systems, which adapt to evolving threats. By assimilating patterns from large sets of trace data, neural networks excel in anomaly detection, allowing the identification of suspicious activities. Furthermore, they could be used in threat classification to select the most optimal countermeasures to be applied.}

\begin{figure}
    \centering
    \includegraphics[scale=0.9, width=\columnwidth]{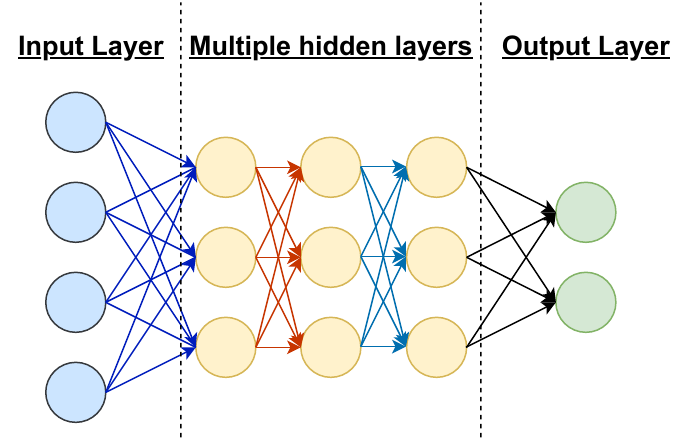}
    \caption{\change{Graphical Neural Network example}{Graphical representation of the architecture of a standard Neural Network.}}
    \label{fig:NeuralNetworkImage}
\end{figure}

    \item \textbf{A* algorithm}: A search algorithm used in AI to find the shortest path or optimal route between an initial node and a target node in a network. This algorithm uses a combination of heuristic search and uniform cost search to determine the most efficient path~\cite{Aalgorithm}. In a reactive system it could be used to find optimal paths in order to select countermeasures based on defined costs. \addtxt{Offering a powerful path finding, efficiently navigating through complex networks to identify the most effective countermeasure selection strategy.} \figurename~\ref{fig:Aestrella} shows a graphical example of the search for an optimal path to solve a problem. In a threat response framework, this algorithm could be used to analyze network or countermeasures graphs, enhancing the system capability to calculate an optimal path.

\begin{figure}
    \centering
    \includegraphics[width=\columnwidth]{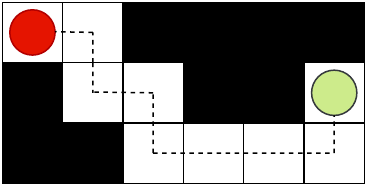}
    \caption{\change{Graphical A* algorithm example}{Graphical A* algorithm example of a path finding solution.}}
    \label{fig:Aestrella}
\end{figure}
    
    \item \textbf{Genetic Algorithms}: Optimization and search techniques based on the theory of biological evolution~\cite{geneticAlgorithm}. These algorithms are used to find approximate solutions to optimization problems. The process resembles the evolutionary process, where candidate solutions are combined, mutated, and selected to form new solutions. Using concepts such as natural selection, crossover, and mutation to improve the solutions. In threat response systems can be used, for example, by evolving a population of potential feature countermeasures in helping to identify the optimal set of security measures. \addtxt{Genetic algorithms can develop a wide range of countermeasure strategies, from access control policies to intrusion detection rules, providing comprehensive protection against cyber threats.}

\end{itemize}

\subsection{Strategy Objectives}

The ``Strategy Objectives'' feature has been selected to clarify the overall objective or strategy behind each chosen methodology. In other words, it could also be defined as a challenge. Additionally, it provides a clear statement of the intended outcome. This information is necessary to understand the motivation behind the selection of the process. Furthermore, it faithfully illustrates the subject matter of the research in question. 

In the following, some of the examples of the values of this feature are further detailed:

\begin{itemize}
    \item \textbf{Minimizing risk}: This objective reflects the intention to reduce exposure to potential risks or adverse events. Such an objective indicates a significant concern to manage any uncertainties that may exist and to ensure the stability and reliability of the architecture. \addtxt{To achieve this goal, organizations must implement practical functions within a structured risk assessment process. This process begins with the identification and classification of all critical information assets. Potential risks are then identified, using tools such as threat intelligence platforms. Vulnerabilities are analyzed to gauge their exploitability and potential impact. Risk is quantified by estimating the probability of each threat occurring and multiplying it by its potential impact. This method allows a quantitative assessment of each threat, which helps to prioritize risks. }

    \item \textbf{Threat detection}: This objective represents the need to identify potential threats. It emphasizes the importance of proactive measures and early detection of possible vulnerabilities. It is the objective presented in the IDS. \addtxt{Threat detection is equivalent to having a vigilant security guard constantly monitoring the premises for any signs of suspicious activity. It involves the deployment of sophisticated tools and techniques that act as digital sentries scanning networks and systems for any anomalous behavior or potential security breaches. If it detects an unauthorized attempt to access sensitive data or unusual patterns indicative of a cyber-attack, it immediately triggers an alert.}
    
    \item \textbf{Minimizing risk and cost}: The combination of these objectives means a dual focus on cost-effectiveness and risk mitigation. It implies a strategy aimed at balancing economic considerations with security concerns. \addtxt{This approach may include conducting comprehensive risk assessments to identify potential vulnerabilities and threats, followed by prioritizing mitigation efforts based on their cost-effectiveness and impact reduction. Organizations, for example, could invest in robust cybersecurity plans that deliver a high return on investment by effectively mitigating common threats while maintaining manageable costs.}
    \item \textbf{Countermeasure standard proposal}: This objective indicates the intention to develop and propose standardized countermeasures. At the same time, it suggests a proactive role in the configuration of best practices and defensive strategies in the field. \addtxt{Standardized approaches are of particular importance, as they offer the advantage of unification. By implementing a standard, information can be shared more directly, easily, and quickly between different response teams. In this way, it not only facilitates communication between teams but also streamlines the process of incident coordination and response, improving the efficiency and effectiveness of security management.}
\end{itemize}

\subsection{Measurements and results}

The ``Measurements and results" feature attempts to encapsulate the measures or indicators used in the investigations. Such specific metrics are used to evaluate the performance and effectiveness of an approach. Therefore, this feature is very relevant as it represents the criteria that describe the degree of achievement of the intended objectives of the methodology. Moreover, it presents the metrics~ \cite{hoffman2019metrics} used by the authors to determine whether the approach has been effective or not. This feature is very valuable, as it could serve as a basis for future studies to identify the most commonly used measures. The use of the same measures provides a comparison of the methods, i.e. the capabilities of the new proposals could be differentiated and compared. 

Among the range of values, the following can be highlighted:

\begin{itemize}
    \item \textbf{Positive rate}: Often referred to as a true positive rate. Is a performance metric that measures the proportion of true positive predictions out of all actual positive instances~~\cite{metricPTR}. Equation~\ref{equ:positiveRate} shows the mathematical expression of this metric\change{. W}{, w}here True Positives \addtxt{(TP)} represent instances that a model correctly recognizes or predicts as belonging to the attribute being tested. \change{And}{Conversely}, False Negatives \addtxt{(FN)} represent instances of the positive class that were incorrectly identified as negative.
\begin{equation}\label{equ:positiveRate}
TP\_rate = \frac{TP}{TP + FN}
\end{equation}

    \item \textbf{Countermeasure selection speed}: Refers to the efficiency with which appropriate countermeasures or defensive strategies are identified in response to threats. Additionally, it can involve quickly evaluating potential countermeasures to mitigate risks.
\end{itemize}

\subsection{Use Case}

The ``Use Case" feature represents those practical scenarios in which the proposed system can be deployed. That is, those real-world contexts or simulated environments in which the proposal might be effectively applied. Such an attribute provides a necessary overview of how the approach can be applied in a variety of contexts. In this way, interested actors can determine whether the approach in the analyzed papers is suitable for their needs. Additionally, this value is necessary for specific individual requirements, thereby ensuring that the application of the solution will be effective in a particular context. 

The range of values is wide, although the following values can be highlighted:

\begin{itemize}
    \item \textbf{Multiple}: This value implicates that the proposed system is flexible and adaptable. Furthermore, it implies that the proposed system is beneficial in a wide range of contexts, resulting in versatility and ease of applicability. 
    \item \textbf{Critical Infrastructure}: This value is applicable in the field of critical infrastructures. Critical infrastructures refer to the fundamental systems, assets, facilities, and networks that are vital for the functioning of a society and its economy~\cite{criticalInfrastructure1,criticalInfrastructure2}. Such a value implies the relevance and effectiveness of the proposed system in safeguarding and optimizing the essential components of the infrastructure.
    \item \textbf{Internet of Things (IoT)}: This value emphasizes the adaptability and functionality of a proposed system into the interconnected IoT ecosystem. IoT refers to a vast network of interconnected devices, objects, and systems that can communicate and exchange data with each other over the internet~\cite{IoT,IoTpaper}.
\end{itemize}

\subsection{Features Summary}\label{featuresSummary}

\begin{figure*}[ht!]
    \centering
    \includegraphics[width=0.8\textwidth]{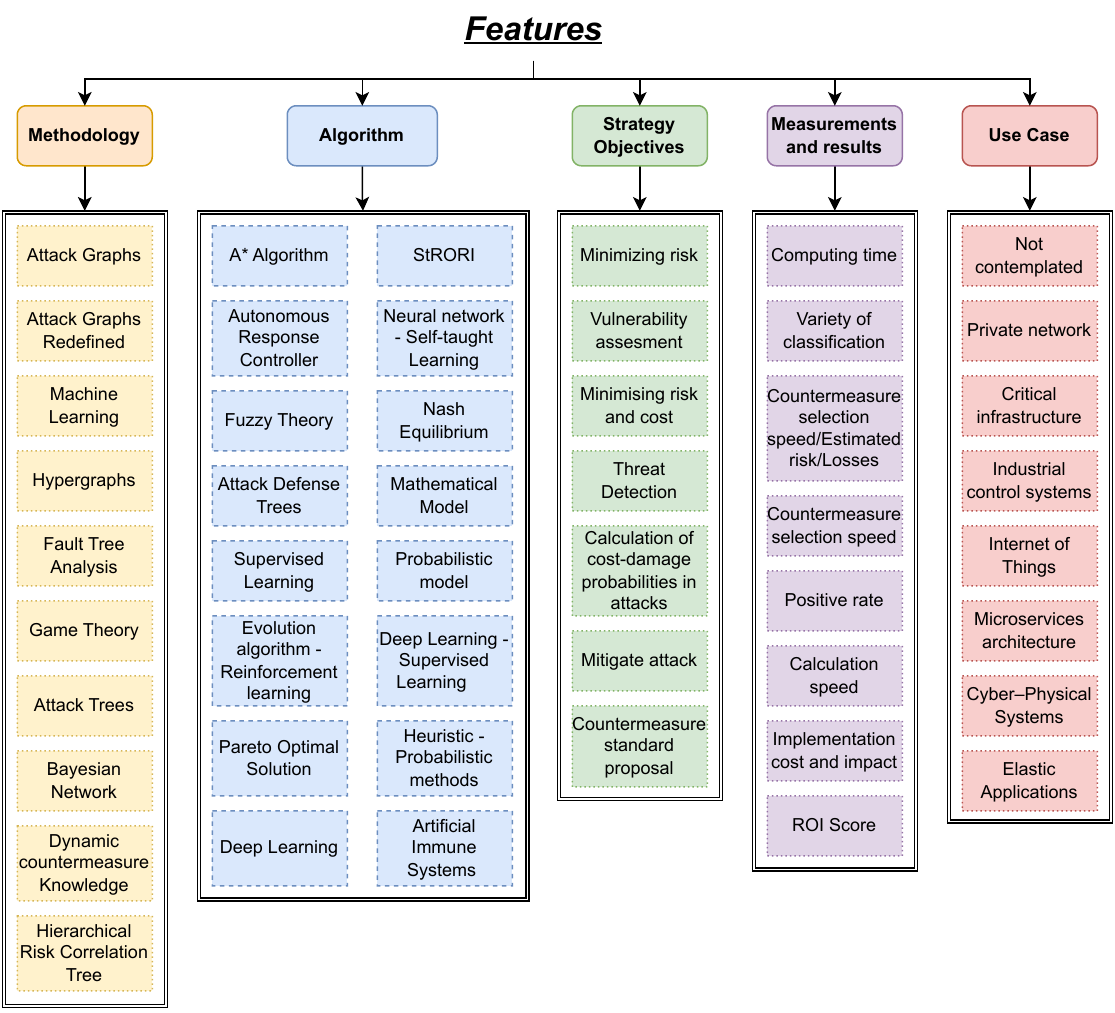}
    \caption{Identified comparison features and values within the selected works}
    \label{fig:featureImage}
\end{figure*}

To conclude, with the intention of facilitating the reading of the survey, a graphic representation of the values identified in each of the features is provided. These values are represented in the \figurename~\ref{fig:featureImage}. In particular, the figure illustrates the five selected features, for which different values are listed:

\begin{itemize}
    \item \textbf{Methodology} where ten values are identified: Attack Graphs, Attack Graphs Redefined, Machine Learning, Hypergraphs, Fault Tree Analysis, Game Theory, Attack Trees, Bayesian Network, Dynamic Countermeasure Knowledge, and Hierarchical Risk Correlation Tree.
    
    \item \textbf{Algorithm} where sixteen values are distinguished: A* Algorithm, Autonomous Response Controller, Fuzzy Theory, Attack Defense Trees, Supervised Learning, Evolution algorithm - Reinforcement Learning, Pareto Optimal Solution, Deep Learning, StRORI, Neural Network - Self-taught Learning, Nash Equilibrium, Mathematical Model, Probabilistic Model, Deep Learning- Supervised Learning, Heuristic - Probabilistic methods, and Artificial Immune Systems.
   
    \item \textbf{Strategy Objectives} where seven values are recognized: Minimizing risk, Vulnerability assessment, Minimizing risk and cost, Threat Detection, Calculation of cost-damage probabilities in attacks, Mitigate attack and Countermeasure standard proposal.
    
    \item \textbf{Measurements and results} where eight values are identified: Computing time, Variety of classification, Countermeasure selection speed/Estimated risk/Losses, Countermeasure selection speed, Positive rate, Implementation cost and impact, Calculation speed and ROI Score.
    
    \item \textbf{Use Case} where eight values are recognized: Not contemplated, Private network, Critical infrastructure, Industrial control systems, Internet of Things, Microservices architecture, Cyber–Physical Systems, and Elastic Applications.
    
\end{itemize}

\section{Survey}\label{Section:Survey}

This section introduces a survey from which the mentioned features have been derived. The purpose is to conduct a thorough comparison among the different papers, to identify existing gaps, so that a unified direction for future research efforts can be proposed in the following sections. By examining and analyzing these studies individually, the aim is to consolidate the most promising approaches and establish a common framework that can guide future work in this field.

\change{For each of the papers a description of the research proposal is conducted. Moreover, an analysis is provided highlighting the advantages and/or disadvantages of each of the approaches.}{For each of the papers a description of the research proposal is conducted. Moreover, an analysis and conclusions of the innovations are provided below highlighting the advantages and/or disadvantages of each of the approaches.} Furthermore, this analysis is based on the features selected in Section~\ref{Features}, ensuring that each paper will be discussed under the same conditions in terms of the proposed features. 

\subsection{\textit{Cybersecurity risk analysis model using fault tree analysis and fuzzy decision theory~\cite{id4}}}
In~\cite{id4}, the authors present a model that integrates Fault Tree Analysis (FTA)~\cite{faultree}, decision theory~\cite{decisionMaking}, and fuzzy theory~\cite{fuzzysetIA} to address two key objectives. FTA provides a structured and systematic way to identify and evaluate potential system failures by modeling the relationships between various components or events within a system. Decision theory is a field that encompasses various methods and tools for making rational and informed decisions. Fuzzy theory is a mathematical framework that allows for the representation of uncertainty and vagueness in a systematic manner. \rmvtxt{On the one hand, the model allows to identify the underlying causes of failures in cyber-attack prevention. On the other hand, it enables to assess the vulnerability of a given system.}

\rmvtxt{Initially, the FTA methodology is employed to establish a failure event associated with a potential cyber-attack and systematically trace its underlying influential factors.} 
Particularly, the primary objective of FTA is to identify the minimal cut set, which represents the minimum combination of basic events leading to the occurrence of the top event, for example the combination: ``CyberAttack - Fault Information System -Unauthorized Access". By analyzing these cut sets, appropriate actions can be prioritized and implemented to prevent the manifestation of the top event and identify vulnerabilities within the system. To comprehensively assess the complexity, magnitude, and impact of a cyber-attack, this research introduces the utilization of fuzzy theory and decision-making techniques under conditions of uncertainty. 

The proposed model is based on five phases: I) expert identification, II) understanding the causes of possible attack scenarios, III) the definition of criteria, IV) fuzzy assessment of potential accidents, and finally, V) aggregation and ordering. In the first phase, a group of experienced persons is selected to maximize the decision-making process and that can be able to identify vulnerabilities in an organization as well as potential accidents within the chance of occurrence of these elements. In the second phase the FTA technique is proposed. The idea is to show a tree representing the failure events, which can represent an outcome of the sequence of the initiation of these events. Particularly, the creation of the FTA procedure consists of four steps:

\begin{enumerate}
  \item Define the system of interest regarding cyber-attacks and establish the initial causes of failure in the security system.
  \item Define the main event of the analysis and specify the problem of interest to be addressed.
  \item The treetop structure is defined, identifying the events and conditions that directly contribute to the top event.
  \item Finally, each branch is explored in detail, recognizing the events and conditions leading to each intermediate event.
\end{enumerate}

The third phase, \rmvtxt{i.e., the definition of criteria,} focuses on addressing the uncertainties involved in risk analysis in the context of cybersecurity. The paper proposes the use of models presented in~\cite{EKEL2008501} and emphasizes the need to consider multiple criteria for evaluating the consequences of a cybersecurity solution. Specifically, the two criteria used in this paper are financial losses and time for restoration. In this sense, the importance of considering cost implications and the ability to repair or replace damaged services quickly is highlighted. The fourth phase focuses on the fuzzy assessment of potential accidents. During this stage, managerial experts evaluate alternative options using the identified criteria. Finally, the last phase involves aggregating all of the criteria used and ordering the alternatives, according to the magnitude of their consequences and using custom equations. These equations are designed for representing maximum risk using an estimated consequence of actions under nature states. The purpose of this representation is used to adopt countermeasures that minimize the risk identified. 

In this paper, an FTA-based approach is introduced, serving the purpose of vulnerability assessment within a system. The use of fuzzy theory is presented as an algorithm for identifying failures in the prevention of cyber-attacks. The methodology aims to minimize risk through the use of customized equations and the evaluation of the different alternatives and states that arise as consequences in the fault trees. A notable concern relates to specific limitations that can affect a reactive system. In the initial phase, where a team of experts manually checks the vulnerabilities of a system, this process can be seen as a disadvantage. To this regard, one can say that the end goal is to develop an autonomous system that can automatically perform these checks. In this respect, it can be said that the ultimate goal is to develop an autonomous system that can automatically perform these checks. This is due to the fact that the objective is to use AI to improve security systems. Among various improvements, one could highlight, in this case, greater efficiency and speed in the identification of vulnerabilities. That being said, the lack of use of AI in this proposal marks another disadvantage. Another drawback detected is the design of the proposal, where the focus is not on the creation of an IRS, but rather on the assessment of vulnerabilities. Nevertheless, some perspectives or proposals that may be relevant to the design of a threat response system can be recognize. For example, the use of FTA can be considered as an advantage as well as for finding vulnerabilities in systems. Especially because of how security incidents are represented and the way in which the occurrence of events can be identified. 

\subsection{\textit{AI- and Metrics-Based Vulnerability-Centric Cyber Security Assessment and Countermeasure Selection: An Artificial Intelligence Approach~\cite{id5}}}

In~\cite{id5}, a set of methods and techniques is proposed for analytical processing of cybersecurity events and information in which the end purpose includes the selection of countermeasures.  The methodology follows an AG-based approach. 

The initial focus of the research is on the data representation used for the countermeasure selection in the cybersecurity context. To achieve this objective, the authors define specific information objects that serve as input for the process. These objects are vulnerabilities, products, weaknesses, exploits, attack patterns, configurations, and countermeasures. The relationships between these objects are established through the analysis of open security databases such as Snyk~\cite{databaseVul} or Mend~\cite{databaseVul2}. In this manner, events can be associated with these objects to enhance their representation and enable more accurate utilization of the information. The data collected by the system can be categorized into two types: \change{input data and output data. Input data encompasses information obtained from external sources, in particular vulnerabilities, code or architecture weaknesses, and attack patterns. It also includes data generated through network analysis, such as network events, configurations, countermeasures, and security policies. In contrast, output data refers to the information generated as a result of attack modeling and analysis, including, AGs, security metrics, impact assessment, selected countermeasures, and identified weaknesses. }{input and output. Input data includes external information such as vulnerabilities, code weaknesses, attack patterns, and network analysis results, while output data comprises the results of attack modeling and analysis, including AGs, security metrics, impact assessment, chosen countermeasures, and identified weaknesses. }The system utilizes both input and output data to enhance its understanding of the security landscape and enable effective decision-making.

\change{Following the data representation, the research introduces the proposed model for analyzing attacks. This model adheres to the subsequent procedures: firstly, an AG is constructed by considering existing and potential vulnerabilities. Next, the actions executed by malicious actors are identified using logs and alerts. Such procedure leads to the generation of an attack subgraph that represents the sequence of possible actions performed by the malicious actor. Furthermore, the model includes the incorporation of potential attack scenarios and corresponding countermeasures, as well as the calculation of security metrics for risk analysis. The resulting information is presented as events and data, enabling the selection of appropriate countermeasures and responses.}{After introducing the data representation, the research presents the proposed model for analyzing attacks. This model follows several steps: i) constructing an AG based on existing and potential vulnerabilities, ii) identifying malicious actions using logs and alerts, iii) generating an attack subgraph to depict the sequence of actions by the attacker, iv) integrating potential attack scenarios and countermeasures, v) calculating security metrics for risk analysis, and vi) presenting the resulting information for selecting appropriate responses.}

The central metric employed is the risk, which is determined by combining the impact of an attack with its probability. To calculate this risk, a Bayesian approach is adopted, utilizing AGs that incorporate probabilities at local, conditional, and unconditional nodes. Each attack action corresponds to the exploitation of a vulnerability, with the Common Vulnerability Scoring System CVSS~\cite{cvss} values for the vulnerabilities being employed in the calculations. This methodology allows for a comprehensive assessment of the risk associated with different attack scenarios, considering both the potential impact and the likelihood of their occurrence.

The paper introduces two countermeasure selection approaches: a static approach and a dynamic approach. The static approach focuses on enhancing system security during design, deployment, and operation by selecting security measures. In contrast, the dynamic approach, used during system operation, monitors events, recalculates risks when necessary, and triggers countermeasures when the risk surpasses predefined thresholds. This dynamic mode ensures real-time responses to evolving threats and prevents attacks. 
 
Within this paper, a methodology is detailed, based on AGs, employing mathematical models as algorithms for countermeasure selection. With this approach, the aim is to mitigate risks within a private network. Overall, this article offers several ideas that can be useful within the reaction ecosystem. For starters, the use of two countermeasure selection modes in this approach is highly interesting as it effectively addresses various aspects of the system's operation. This technique can provide comprehensive coverage and ensure that all necessary measures are taken into account. To continue, the detailed representation of information regarding different events recorded within the system proves to be indispensable for the identification and representation of security incidents. No significant concerns have been identified, apart from the lack of standardization in the design of countermeasures, which could have been of interest. 

\subsection{\textit{Heuristic Approach Towards Countermeasure Selection using Attack Graphs~\cite{id1}}}

In~\cite{id1}, an approach of heuristic search is followed to select the optimal countermeasure deployment under a given budget constraint. The approach implements AGs to model the risk present in the system. In order to do this, the proposed system considers vulnerability exploitation consequences and examines potential attacker lateral movements. Additionally, it represents the network's physical topology as well as the vulnerabilities of the network protocols to minimize the system risk within the budgetary limitations. 

The solution relies on the shortest path problem where the A* heuristic algorithm~\cite{Aalgorithm} is used to find the most optimal set of countermeasures efficiently. The novelty of the article relies upon using a particular risk equation, which indicates the presented risk of exploiting an asset provided by CVSS scores, together with the associated countermeasure, thus allowing the evaluation of the risk of the system without regenerating the AG, saving computational time. MulVAL~\cite{MulVal} is used as a security analyze based on logic, serving as a tool for the analysis of AGs. This tool uses the Prolog~\cite{prolog} language to represent the pre and post-conditions for exploiting a vulnerability and then identify the necessary countermeasure. \change{The authors define countermeasures as security products with two attributes, on the one hand, the defense mechanism they implement, such as a firewall or Antivirus, and, on the other hand, the set of mitigation actions, for example disabling/enabling ports.}{The authors define countermeasures as security products characterized by their implemented defense mechanisms, such as firewalls or antivirus software, and the associated mitigation actions, like disabling or enabling ports. } \change{The proposed system works as follows. As a starting point, a risk assessment is carried out using the AGs. Subsequently, relevant countermeasures are identified by matching them with mitigation actions at the nodes of the AG, and identifying where the actions would be executed in the network. As a next step, the risk calculation is performed, using the risk equations that come from the preprocessing phase of the AGs. Once the risk has been calculated, the A* algorithm is used to obtain the shortest path, which in this case corresponds to the one with the lowest risk, considering the cost, and thus identifying an optimal countermeasure plan. Finally, an evaluation of the proposal is carried out. For this purpose, costs are assigned as an effort for the deployment of each countermeasure. The algorithm will use a general budget and will reduce the risk, reducing the budget as an expense after the deployment of the corresponding countermeasure. It can be seen that the risk of the system decreases as the budget increases, but also the execution time of the algorithm increases because additional countermeasures must be considered, which might incur higher costs.}{The proposed system begins with a risk assessment using AGs. Relevant countermeasures are then identified by matching them with mitigation actions at AG nodes and determining their network execution points. Risk calculation follows using equations derived from AG preprocessing. The A* algorithm is employed to find the shortest path, minimizing risk while considering cost, and yielding an optimal countermeasure plan. Evaluation involves assigning costs for countermeasure deployment, with the algorithm adjusting the budget based on risk reduction. As the budget increases, system risk decreases, but algorithm execution time may rise due to consideration of additional, potentially costlier countermeasures.}

As mentioned before, this paper presents a methodology based on AGs for the purpose of risk minimization of a system. For this purpose, an A* algorithm is used for the selection of countermeasures. Analyzing the paper, a disadvantage identified concerns the limitations that conflict with the policies of the organization in which the system is implemented. In other words, the algorithm does not take into account certain limitations when selecting countermeasures. That is, in the risk graph shown in the paper, the higher the budget, the more the risk is reduced, even to 0\%. This is however not useful in the real world, since the algorithm could propose countermeasures that disable relevant services or communications. Nevertheless, the use of MulVAL for AG security analysis is an interesting point. Furthermore, it alludes to establishing a maximum cost, on which the heuristic-based algorithm is built, which increases the realism of the design. Therefore, it is easy to spot that this is a very interesting proposal that, once the possibility of taking into account certain requirements is added, can be very relevant. 

\subsection{\textit{Introducing Deep Learning Self-Adaptive Misuse Network Intrusion Detection Systems~\cite{id2}}}

Due to the large number of interconnected devices that may exist in a network, the ability to handle their capabilities is typically beyond human limits. In the same way, IDSs are unable to automatically adapt to the many changes that occur in their environment, which reduces their efficiency. For example, by inserting a new device into the network it may generate services or communications or even vulnerabilities never seen in the network before. To solve the aforementioned challenges, in~\cite{id2}, Deep Learning (DL) is proposed to identify unknown threats in unknown environments. In particular, a self-taught learning methodology is used, an ML methodology that can train itself using unlabeled data to improve a supervised classification problem. This methodology is integrated within the MAPE-K structure~\cite{MAPEK}, which is based on the steps Monitor (M), Analyze (A), Plan (P), and Execute(E) while using a Knowledge (K) base as support. The proposed system is based on environmental states and network flows. \rmvtxt{Authors claim that network} \addtxt{Network} changes \change{can be}{are} seen as potential incidents that may contain attacks or vulnerabilities. For this purpose, network flows are used, which, by possessing  relevant information, can identify unknown and unseen network flows that may indicate the presence of previously unknown attacks. 

In the monitor phase, the sensors of the system are coordinated to acquire the basic knowledge of the network, as well as network mappers~\cite{netmapper} generate a network inventory. \change{The idea is to identify}{Identifying} the network characteristics alteration events that require activating the IDS. In parallel, network sniffers are used to capture network flows through the detection engine of the IDS. In this phase, network traffic is stored in the Knowledge base for later phases. In the Analyze phase, an analyzer leverages network audit tools to generate structured network flows from network traffic that can bring various features. These features are fed into the IDS engine, which will in turn use a supervised model for detection purposes and, on the other hand, it will be used as unlabeled data for the self-taught learning adaptive process. 

Up to this point, the Monitor and the Analyzer identified environmental changes in the network, while the Knowledge component consolidated the network flows generated at the time the changes occurred. At this moment, the IDS may face unknown network traffic, which is why the planning phase undertakes to utilize unsupervised feature learning techniques. This technique is an NN that applies back-propagation called sparse auto-encoder~\cite{makhzani2014ksparse} to learn sparse representations of the input data. The output of the planning phase is a feedforward auto-encoder that is used to reconstruct the initial set of labeled data into a new training dataset for a supervised learning algorithm. After training the new model, the old model, which due to environmental changes had started to have efficiency problems, can be replaced. 

For the purpose of detecting threats within a system, this paper introduces a methodology relying on ML. The detection is facilitated by employing an efficient NN that uses self-taught learning. This approach can offer some notable advantages. First and foremost, the use of ML provides the system with the ability to adapt and learn from new data, while the employment of a self-taught NN ensures that the system can autonomously discover patterns and anomalies in the data. However, one drawback identified in this paper has to do with the strategy objective. Where the purpose is not based on the creation of an IRS, but rather on the identification of potential threats. However, this proposal shows a very interesting idea that can be very useful when detecting unknown threats based on changes in the state of the network. Using the states of the network, it might be possible to model a system that could be under attack and generate, through AI, different situations and identify each of the countermeasures that could be applied, anticipating possible results of the application of such measures. As an additional advantage, the usage of MAPE-K methodology was presented, as it may be relevant when creating an IRS using AI that can retrofit its intelligence while running. For these reasons, the ideas and perspective of the proposed methodology could be used in the design of countermeasures and their impact-based adjustment to improve the response to intrusions.

\subsection{\textit{A Dynamic Decision-Making Approach for Intrusion Response in Industrial Control Systems~\cite{id24}}}

The paper~\cite{id24} introduces a decision-making approach for addressing intrusion response in Industrial Control Systems (ICS). The primary objective of this approach is to promptly determine the optimal security strategy against cyber-attacks, safeguarding the most dangerous attack paths and responding to functional failures efficiently. Unlike conventional approaches, this decision-making model focuses on both the cyber and physical domains, enabling a comprehensive analysis of attack propagation and potential vulnerabilities within the ICS environment.

To achieve its goals, the approach formulates a set of measures that cover various aspects of security in-depth, including defense and recovery measures. These candidate security measures are carefully designed based on a thorough analysis of attack propagation, ensuring the completeness of the candidate security strategy space. Additionally, the approach embraces a dynamic and adaptive approach to intrusion response, enabling it to address evolving and sophisticated cyber-attacks effectively.

\change{A crucial aspect of this decision-making approach is the use of a multilevel Bayesian network model, which offers a profound understanding of how cyber-attacks propagate from the cyber domain to the physical domain within the ICS. This modeling allows for a comprehensive representation of the implications of cyber-attacks and their potential impacts on the entire system. By leveraging this multilevel Bayesian network model, the approach can accurately map detected evidence from real-time IDS, thereby constructing a comprehensive candidate security strategy space based on attack surfaces and functional failures.}{A key aspect of this approach is the use of a multilevel Bayesian network model, providing insight into how cyber-attacks spread from the cyber to the physical domain within the ICS. This modeling enables a comprehensive representation of the consequences of cyber-attacks on the entire system. By utilizing this model, the approach accurately maps evidence from real-time IDS to construct a comprehensive security strategy space based on attack surfaces and functional failures. }

To select the most effective security strategy, the approach employs Multiobjective Optimization (MOO), aiming to maximize the objective vector comprising security benefit, system benefit, and state benefit. Balancing these three benefits is crucial to ensure the optimal protection ability of the chosen strategy while considering the system's requirements and impacts. The approach uses a distance-based evaluation method to prioritize the Pareto optimal solutions within the candidate security strategy space, ultimately selecting the optimal strategy that best aligns with the ideal objective vector.

As already said, to effectively prioritize the solutions and strike a balance between the three benefits, a distance-based evaluation method is designed, comprising three key steps. Step 1: Finding the ideal vector in this step, the ideal vector is identified, representing the perfect solution in a multidimensional space. Step 2: Normalizing each vector in the Pareto Front, this refers to the set of non-dominated solutions representing the best trade-off solutions between the three objectives. In this step, each vector in the Pareto front is normalized to bring it into the same scale. Step 3: Evaluating the distance of each normalized vector after normalization, the distance-based evaluation method computes the distance between each normalized vector and the ideal vector. The distance represents how close a particular solution is to the ideal state in the multidimensional space. Smaller distances indicate a higher level of optimality, suggesting that the solution is closer to the ideal performance across all three objectives.

In conclusion, the presented security decision-making approach for ICS showcases several noteworthy advantages. Its ability to promptly respond to cyber-attacks and ensure a comprehensive candidate security strategy space makes it a valuable tool for bolstering ICS security. The incorporation of fine-tuned candidate security measures through MOO enables a balanced consideration of security benefit, system benefit, and state benefit, effectively optimizing security performance while preserving the overall system's functionality. However, a potential limitation lies in its focus on the most dangerous attacks, possibly overlooking less conspicuous threats that may still pose risks and be exploited by attackers. To further enhance its effectiveness, future developments could explore more inclusive and adaptable strategies to address a broader range of potential attack scenarios and fortify ICS environments against ever-evolving cyber threats. To conclude, the absence of AI in the proposal is noteworthy. The incorporation of AI models has the potential to effectively mitigate adaptability challenges within the strategy. Furthermore, the use of these models offers the advantages of identifying the most dangerous attacks and improving the decision-making process.

\subsection{\textit{Stateful RORI-based countermeasure selection using hypergraphs~\cite{id3}}}

\cite{id3} lies in the development of a dynamic tool based on a Stateful Return on Response Investment (StRORI) index that uses a hypergraph model for evaluating countermeasures based on financial and threat impact assessment functions. The StRORI metric is used to rank countermeasures and select the optimal for reducing the impact of detected attacks. 

Concretely, the Return on Response Investment (RORI) model was originally proposed in~\cite{serviceDependecy} as an index to compare response candidates and determine the optimal response. The authors used an improved version that uses different parameters related to lose, cost, and risk mitigation levels. By leveraging RORI, one can effectively identify the most suitable countermeasures to mitigate pre-determined attack scenarios. The inclusion of ``Stateful" within StRORI introduces a novel approach. That is specifically tailored to accommodate the dynamic nature of the monitoring system. In particular, such an index is able to achieve this operation by maintaining the state of the network at all times so that, at each execution, a unique state-based representation of the system is generated. This unique representation allows for the comparison of system parameters each time the network is evaluated. Besides, the use of the hypergraph is based on the decision to reduce the size of the graphs from the initial AGs, which avoids facing their enormous size that hinders their efficient automatic and manual processing and visual perception. In this sense, authors claim that employing such graphs offers notable benefits, including the capability to incorporate supplementary semantic information within the edges, facilitating the grouping of diverse object types that share associations, such as countermeasures.

In this article, the authors propose two modes of operation: preventive and reactive. In the preventive mode, there is a preventive countermeasure selection implemented for the graph nodes that present unacceptable risk levels from a predefined threshold. The goal is to reduce the risk level with minimum cost. The reactive mode presents a dynamic response when a cyber-attack is detected. The goal is to prevent or stop propagation before any more severe damage is done. These three phases are used: 

\begin{itemize}
  \item \textbf{\textit{Incident mapping}}, where the security incident is mapped into an AG.
  \item \textbf{\textit{Risk calculations}}, where risks are recalculated from previous graphs.
  \item \textbf{\textit{Reactive countermeasure selection}}, where countermeasures are selected.
\end{itemize}

A methodology firmly based on hypergraphs is presented in this paper, addressing the purpose of minimizing the risk and cost within a system. Using an effective StRORI model, optimal countermeasures are selected. It is noteworthy to mention that one disadvantage observed is the absence of AI. Although it is a dynamic tool, no mechanism is used to allow the system to learn and adapt based on the information it gathers. Furthermore, the countermeasures used are not described. Neither a standard for countermeasures is apparently used. Nonetheless, several advantages or ideas can be drawn from the proposal including, for example having two different modes of operation provides a good overall approach to addressing cybersecurity challenges. The preventive mode effectively identifies and reduces risks statically by acting on nodes with unacceptable risk levels, minimizing potential damage and costs. At the same time, the reactive mode responds quickly to detected cyber-attacks, stopping their spread and preventing more serious consequences. The use of an index such as StRORI can be presented as well as an interesting advantage that helps the dynamism of the scenario analysis. In combination with the hypergraph model, this provides an improvement in the accuracy and effectiveness of countermeasure selection.

\subsection{\textit{Exploiting attack–defense trees to find an optimal set of countermeasures~\cite{id9}}} 

In~\cite{id9} it is propose a comprehensive framework that leverages Attack-Defense Trees (ADTrees) in conjunction with Integer Linear Programming (ILP) to identify an optimal set of countermeasures for system security. ADTrees, a variant of AGs, introduce the concept of a defender alongside the attacker, enabling a more natural representation of the game between the two actors. The framework provides a constructive approach to extract all plausible behaviors of both attacker and defender from ADTree models. The nodes' labels in the ADTrees depict the goals of the two actors, allowing for a comprehensive analysis of their interactions and strategies. Drawing on this extracted information, employing ILP to generate a generic solution capable of addressing a wide range of optimization problems related to security.

\change{By formalizing ADTrees using DAG, the ability to reason conveniently about the attacker's actions is gained, which may contribute to multiple distinct attacks, as well as the countermeasures that can mitigate various attack vectors. The framework offers flexibility by formulating a generic solution using ILP, allowing for customization to specific security-related optimization criteria. }{Formalizing ADTrees using DAG enables convenient reasoning about attacker actions, facilitating analysis of multiple attacks and corresponding countermeasures. The framework's flexibility lies in its generic solution using ILP, allowing customization to specific security optimization criteria. }

\change{The proposed approach introduces a series of lemmas and corollaries that form an algorithm for extracting possible rational behaviors from the ADTree model in a security environment. These behaviors are defined as strategies, characterized by sets of basic actions performed by both the attacker and defender. This algorithm allows for a systematic exploration of the different ways in which the two actors can interact within the ADTree framework.}{By formalizing ADTrees using DAG, the ability to reason conveniently about the attacker's actions is gained, which may contribute to multiple distinct attacks, as well as the countermeasures that can mitigate various attack vectors. The framework offers flexibility by formulating a generic solution using ILP, allowing for customization to specific security-related optimization criteria.}

Furthermore, the framework incorporates defense semantics, which captures the rational approaches for the defender to achieve the root goal. The root goal represents the primary objective of the modeled scenario, and the defense semantics provide a comprehensive representation of the various viable paths the defender can undertake to attain this goal. By considering these defense semantics, the framework enhances the understanding of the defender's strategies and aids in the selection of appropriate countermeasures to mitigate potential threats.

The framework capitalizes on defense semantics to create a flexible solution for handling optimization problems expressed as ILP formulations. Furthermore, the framework incorporates an additional constraint that considers the cost associated with each action. By integrating defense semantics and cost considerations, the framework enables the efficient resolution of optimization problems by formulating and solving ILP models.

The modeling methodology and defense semantics are automated using an open-source tool called Optimal Strategies Extractor for Attack-Defense trees (OSEAD). This software offers a comprehensive solution for solving various problems, including the determination of a set of countermeasures that effectively neutralize the most cost-efficient strategies employed by the attacker. To utilize OSEAD, users simply need to provide an ADTree model, along with the budget and cost information for each action. By leveraging this tool, users can streamline the process of identifying optimal countermeasures while considering budgetary constraints. OSEAD empowers security practitioners to make informed decisions regarding the allocation of resources, ensuring efficient defense strategies are implemented to mitigate risks effectively.

Within this paper, a methodology is detailed, based on AGs, to serve the purpose of minimizing risk within a system. Employing ADTree appropriate countermeasures are selected. The first comment about this paper has to do with the lack of use of AI. However, ILP has been used together with DAG and ADTrees.  It is important to be aware of the fact that the use of ADTrees could provide scalability limitations that could be seen as a constraint in scenarios with many assets~~\cite{representationModels}. The automation of the modeling methodology through the proposed tool is a noteworthy aspect. Another advantage of using the OSEAD tool relates to the estimation of the cost of each countermeasure. This provides organizations using the framework with the ability to limit countermeasures according to the budget. To further enhance the proposal, the integration of AI could play a pivotal role in tasks such as automating the creation of ADTree models for different network configurations. Additionally, incorporating dynamic reformulation capabilities to account for changes resulting from attacks or the implementation of countermeasures would enhance the overall adaptability and robustness of the approach. It is therefore considered as an interesting research, which may contain several relevant concepts, yet it might become more autonomous through the use of AI.

\subsection{\textit{Bayesian Decision Network-Based Security Risk Management Framework~\cite{id17}}}

In~\cite{id17}, an integrated framework for network security risk management is presented, leveraging a powerful probabilistic graphical model known as the Bayesian Decision Network (BDN). This novel approach utilizes BDN to effectively model and manage security risks by incorporating crucial information, such as vulnerability data, risk-reducing countermeasures, and the potential effects of their implementation on vulnerabilities. To enhance the precision of the risk assessment process, the proposed framework takes into account various factors, including inherent characteristics of vulnerabilities, temporal dynamics, and environmental influences. 

The risk management framework comprises three main phases: risk assessment, risk mitigation, and risk validation and monitoring. 

The first phase, risk assessment, focuses on modeling network attacks using AG and calculating the temporal probability of successful vulnerability exploitation. This phase is divided into four steps: modeling network attacks, probability calculation, Bayesian Attack Graph (BAG) construction, and impact calculation. \change{The initial step of the proposed approach necessitates gathering information on existing vulnerabilities, network topology, and host connectivity to generate an AG. To facilitate this process, the authors recommend utilizing tools like MulVAL for AG generation}{The initial step involves gathering information on existing vulnerabilities, network topology, and host connectivity to generate an AG, possibly utilizing tools like MulVAL. Transitioning to the third step, the AG is transformed into a BAG by introducing Conditional Probability Tables (CPTs) to each of its nodes, ensuring that the AG remains an acyclic graph. }. \change{Moving on to the second phase, probability calculation, security administrators must incorporate the exploitation probability for each vulnerability into its respective nodes using CVSS metrics. Next, the AG is transformed into a BAG with Conditional Probability Tables (CPTs) to ensure acyclicity. }{Security administrators then incorporate exploitation probabilities into each vulnerability node using CVSS metrics. } Finally, the impact of vulnerabilities is computed, considering metrics such as confidentiality requirement, integrity requirement, and availability requirement. \rmvtxt{These calculations provide crucial insights into the potential consequences of exploiting vulnerabilities within the network's security framework.}

In the risk mitigation phase, the framework identifies and assesses potential countermeasures to mitigate vulnerability exploitability. This phase is divided again into four steps: countermeasure analysis, BDN construction, filling utility tables, and cost-benefit analysis. The risk management framework consists of four concise steps. First, in the countermeasure analysis, security experts identify all relevant countermeasures to reduce risk and implement them on vulnerabilities. This hinders attackers' efforts and reduces the vulnerability's exploitability. Next, the BDN construction employs a model compatible with AG and BAG, enhancing the risk mitigation process. Decision and utility nodes are added to convert BAG models to BDN models for networks. The third step involves filling utility tables where the security administrator chooses to implement or omit specific countermeasures. Lastly, the Cost-Benefit Analysis identifies the optimal subset(s) of countermeasures to maintain an acceptable security level within the network.

The third phase, risk validation, and monitoring, involves continuous tracking of changes in network states to ensure that the residual risk aligns with the desired risk level. This ongoing monitoring ensures the framework's effectiveness in managing risks over time, allowing for timely adjustments and improvements as needed.

The paper introduces an approach based on a Bayesian network methodology. For the purpose of minimizing risk in a system, a probabilistic model is used as algorithm. The framework, utilizing BDNs and BAG, effectively minimize security risks by incorporating vital information about vulnerabilities, risk-reducing countermeasures, and their potential impacts. The proposed framework demonstrates promising ideas in preventive security management. Such as the use of CVSS metrics or the CPTs for the transition to BAG. Nevertheless it has some limitations worth considering. One key limitation is its dependence on security administrators for decision-making during mitigation phase. This is a disadvantage of the model, since the aim of a reactive autonomous system is full task automation and non-reliance on human interaction. In addition, the document focuses mainly on preventive security management, while offering very limited insight into the modelling of reactive countermeasures. This aspect may limit its ability to respond and mitigate incidents in real time.

\subsection{\textit{Dynamic Countermeasure Knowledge for Intrusion Response Systems~\cite{id19}}}

In~\cite{id19} the authors present a countermeasure standard that enables the sharing of countermeasure intelligence and the automated adoption and execution of countermeasures by an IRS. The main goal is to develop a dynamic countermeasure knowledge system that can effectively prevent attacks, especially those that a static predefined countermeasure approach would not be able to stop.

The paper defines a countermeasure as a direct attempt to halt or mitigate the damage caused by an ongoing, predicted, or executed attack. Currently, building a collection of countermeasures for an IRS involves manual generation by a security analyst or the IRS developer. However, with the aim of achieving a fully automated intrusion response process, it becomes essential to remove the manual aspect of generating and maintaining the countermeasure collection.

As part of the development of dynamic countermeasure knowledgebase, the paper proposes a format for countermeasures that enables automated execution. This format, called Common Countermeasure Format (CCF), is presented as an XML structure, building on the CRE entry proposed by NIST. CCF incorporates attributes that facilitate machine-readable atomic actions, aiming to establish a universal standard format for countermeasures, similar to the functionality of the Common Vulnerabilities and Exposures (CVE)~\cite{CVE} format. The primary objective of CCF is to enable countermeasure knowledge sharing across various domains and to provide a means for future IRS to dynamically adjust their collection of countermeasures in response to the vulnerabilities present in the network.

CCF is proposed as an upgrade for the creation of an online repository of countermeasures, regularly updated to reflect the latest developments in cybersecurity. New entries in this repository will follow a similar approval process as CVEs. This repository's availability will allow IRS implementations to periodically query and access the latest relevant countermeasures, much like vulnerability scanners use National Vulnerability Database (NVD) for up-to-date vulnerability information.

This paper focuses on a countermeasure standard proposal. For this, the authors propose a dynamic countermeasure knowledge model. On a slightly negative note, the document focuses only on outlining the format of countermeasures. The development of a countermeasure selection system is left aside. However, this approach is very interesting, as a common countermeasure format is necessary to unify efforts and develop autonomous reaction systems. One of the advantages of this approach over the countermeasure standard is a dynamic countermeasure knowledge base. By adding dynamism to countermeasures it is easier for organizations to maintain a good number of countermeasures. That is, as new types of malware emerge, or as attacks have evolved, countermeasures will also evolve in the same way. And, since it is a shared database, all systems with such a knowledge base will benefit from updates. The only drawback that can be identified with this type of distributed database is the maintenance. That is, who would be in charge of monitoring the countermeasures that are included, as well as the quality of these countermeasures? It is important to do this because if access is freely open, malicious actors could modify countermeasures to be seen as valid while hiding vulnerabilities in the countermeasures.

\subsection{\textit{A hybrid model-free approach for the near-optimal intrusion response control of non-stationary systems~\cite{id20}}}

In~\cite{id20} the authors present a hybrid model-free IRS based on deep Reinforcement Learning (RL)~\cite{Reinforcementlearning}. The experimental results demonstrate the effectiveness of the proposed IRS, particularly in dealing with non-stationary systems.

The proposed approach involves four distinct phases. Firstly, the system model is designed, to prepare for an algorithm based on planning. In the second phase, a software simulator of the system is constructed based on this model. Next, in the third phase, an RL agent is integrated with the simulator, allowing it to autonomously learn the system model. Finally, in the fourth phase, the RL agent is disconnected from the simulator and connected directly to the actual system. In this real-world setting, the agent can detect deviations from the model and dynamically adapt to the evolving system. The concept of transfer learning underpins this process, enabling the agent to draw upon knowledge from related systems, resulting in reduced training time and more efficient adaptation to new environments. 

The proposed IRS is designed with a stateful model of the system, which empowers it to strategically plan sequences of defense actions upon detecting an attack initiated by an IDS. The underlying system model is based on an Markov Decision Processes (MDP)~\cite{sutton2018reinforcement}, with a primary focus on applications that utilize the microservices architecture style. The core element of the model is the ``ComponentGroup" representing the microservices, composed of various instances. Each component is characterized by multiple state variables, including:

\begin{itemize}
\item Active: This variable determines whether the component is currently turned on or off.
\item Updated: It indicates if the element has been updated to the latest version.
\item New Version Available: This variable specifies whether a new software version is available to update the current one.
\item Corrupted: It indicates whether a component has been compromised.
\item Vulnerable: This variable states whether the current software version of the component is vulnerable to a known attack.

\end{itemize}

By considering these state variables, the IRS can make informed decisions about the appropriate defense actions to be taken in response to detected threats. The utilization of MDP and the incorporation of microservices-oriented components enhance the IRS's ability to effectively respond to dynamic cyber threats.

In order to minimize risk in microservices architectures, this paper proposes a ML methodology. As an algorithm, the research uses an evolution algorithm, such as RL with MDP. However, it is important to note that the proposed approach is exclusively based on the model of the defended system and does not consider the model of the attacker. Consequently, the IRS is purely reactive and relies on an IDS to detect any additional steps of a multi-stage attack. Despite this limitation, the experimental results indicate that the proposed IRS effectively handles non-stationary systems. The hybrid model based on deep RL is an interesting point and the research shows the effectiveness of this type of methodologies. It is worth commenting on the use of the state model of the systems being evaluated. Being able to differentiate the characteristics of each of them in a state is an interesting idea that can help in decision-making and the generation of security measures. To conclude, as a drawback, it is worth noting that the system has only been tested in a laboratory environment, so it may not be as effective as in a realistic environment. 

\subsection{\textit{Autonomous mitigation of cyber risks in the Cyber–Physical Systems~\cite{id21}}}

In~\cite{id21} the authors present a contribution in the field of Cyber-Physical Systems (CPS) security by proposing an innovative approach for autonomous response against cyber-attacks. The authors introduce the Autonomous Response Controller (ARC), which marks a significant step toward achieving efficient and automated intrusion response in CPS environments.

The ARC is designed to operate in a hybrid model-free framework, \change{offering the capability to respond to cyber-attacks with or without human intervention, depending on the criticality of the CPS assets at risk.}{responding to cyber-attacks autonomously or with human intervention based on CPS asset criticality.} \change{The framework incorporates the quantitative Hierarchical Risk Correlation Tree (HRCT), which effectively models attacker paths and quantifies the financial risk that CPS assets may face from cyber-attacks.}{It incorporates a quantitative Hierarchical Risk Correlation Tree (HRCT) to model attacker paths and quantify financial risks to CPS assets.} \rmvtxt{By leveraging the HRCT, the ARC gains a comprehensive understanding of the potential threats and vulnerabilities within the CPS.}

\change{The power systems serve as a prime application domain for the proposed approach, and the ARC exhibits its prowess by delivering a scalable and autonomous response strategy for the CPS. The risk assessment model based on HRCT empowers the ARC to analyze various security enhancement options, evaluating their effectiveness and prioritizing them based on relative benefits and cost justifications.}{In power systems, the ARC delivers a scalable and autonomous response strategy, analyzing security enhancement options and prioritizing them based on benefits and cost.} This dynamic decision-making capability ensures that the ARC can adapt and evolve its responses in real time\addtxt{.} \rmvtxt{, keeping the CPS protected against both known and emerging threats.}

The ARC's adaptability is evident as it overcomes several challenges inherent in today's IDSs. Despite the absence of precise alerts matching successful intrusions, the ARC computes appropriate responses, offering a strong defense mechanism against cyber-attacks. Additionally, the ARC's capability to create long-term response plans showcases its effectiveness in handling complex multi-stage attacks orchestrated with intelligent planning by adversaries. Taking into account specific CPS characteristics and requirements, such as asset criticality, resource availability, and response impact, the ARC dynamically adjusts its strategies to optimize security and minimize potential risks.

A remarkable aspect of the ARC lies in its timely responses against ongoing attacks, facilitated by the low-security state space derived from HRCT. This enables the ARC to swiftly adapt and react to evolving threats, ensuring the CPS remains resilient and safeguarded against potential compromises.

HRCT-based probabilistic risk assessment model provides valuable insights into complex attack scenarios, where multi-stage attacks are viewed as sequences of elementary attacks. The experiments conducted to evaluate HRCT's performance highlight its efficacy in accurately analyzing the risk in CPS networks with a remarkably low error rate. 

In conclusion, the paper introduces a novel approach for enhancing security by minimizing the risk in CPS. The methodology used for this task is based on the use of HRCTs. The core of this proposal is an ARC algorithm. Several advantages are noteworthy to mention. To begin with, a strong aspect of this framework has to do with the responsiveness of the framework. It is a very attractive idea how the framework is designed to intervene against cyber-attacks based on the criticality of a particular asset. This can be very beneficial for those scenarios where highly critical machines are attacked, and the reaction time does not allow for human intervention. Another positive aspect is related to the generation of long-term security strategies. This could be seen as a two-mode system, in which knowledge is generated and used statically, and when an attack is identified, it is acted upon autonomously and quickly. However, there is a lack of standardization or proposals for knowledge generated through an attack. Such knowledge could be relevant to generate an intelligence base to improve reaction systems through improved AI.

\subsection{\textit{A Bio-Inspired Reaction Against Cyberattacks: AIS-Powered Optimal Countermeasures Selection~\cite{id16}}} 

In the paper~\cite{id16} a novel cybersecurity methodology based on Artificial Immune Systems (AIS) is proposed. This methodology leverages the exceptional properties of bio-inspired techniques to select optimal sets of countermeasures. Inspired by biological processes, the methodology incorporates cloning and mutation phases to enhance the quality of the solution in combating cyber threats. By harnessing AIS, this innovative methodology provides a unique and effective approach to selecting countermeasures, thereby improving overall cybersecurity response.

The authors build upon their previous work~\cite{prestandardization}, which proposed a standard representation for countermeasures in the field. This representation serves as a foundation for their research. The significance of considering the trade-off between the effectiveness of countermeasures and their potential negative impact and cost is emphasized. \rmvtxt{The importance of selecting an optimal set of countermeasures that achieves the desired remediation while minimizing any adverse effects is highlighted.}

The methodology incorporates distinct phases to enhance the quality of the solution in combating cyber threats: It can be fired at a preventive phase  (AIS Static reaction) due to its ability to prevent potential threats from happening, or at a reactive stage (AIS Dynamic reaction) since it also is capable of eradicating an ongoing attack and remedying its adverse effects.

Additionally, the paper highlights the valuable features of AIS. These features include uniqueness, where each reaction is tailored to a specific threat similar to the immune system. The distributed reaction and self-regulation aspect eliminate the need for central coordination and control during reactive immuno-operations. Diversification is achieved through clonal selection and hypermutation, constantly generating improved responses to protect system assets under attack. The self-protection characteristic ensures that the reactive immuno-reaction only focuses on the designated assets, minimizing negative impacts. Lastly, the capability of memorization allows the system to optimize responses based on previous reaction information, enhancing future defense mechanisms.

On the one hand, AIS Static reaction is designed to be deployed at a preventive phase, where it effectively prevents potential threats from materializing. By analyzing the system's vulnerabilities and potential attack vectors, the AIS can proactively identify and select appropriate countermeasures to fortify the system's defenses. In this case, the vulnerability represents the antigen, and the set of atomic countermeasures of an asset, are the antibodies that must be selected.

On the other hand, the AIS Dynamic reaction is deployed at a reactive stage, where it demonstrates the ability to eradicate ongoing attacks and remediate their adverse effects. In the event of a detected attack, an intrusion detection message is shaped and sent to the SIEM for further investigation, containing meaningful information about the attack. The AIS quickly responds by selecting and implementing countermeasures to neutralize the threat and mitigate its impact. The method will first generate initial random antibodies (set of atomic countermeasures) and will calculate the affinity. The antibody with the highest affinity will be cloned and mutated and the antibodies with the lowest affinity are removed from the solution space. Within a loop, the best antibody near the acceptable risk level is chosen as a final solution, taking into account the prioritizing of the timing factor.
 
In conclusion, the paper uses ML to address the challenge of minimizing risk in information systems. It will also select the optimal set of countermeasures that effectively eradicate the threat while minimizing negative impacts on system properties. For this, an AIS-based algorithm is used. The results of this paper show that the proposal of linking biology and cybersecurity is a promising concept. In addition to this, one of the advantages is the design of the different modes involved in the methodology. On the one hand, there is a preventive phase that scans the system for potential weaknesses. On the other hand, a reactive phase has been designed, in which ongoing attacks are remediated. This type of system is recurring in previous analyses, and it seems to perform successfully as it divides tasks between the two phases. Furthermore, another significant advantage of this paper is that it attempts to unify the countermeasure format, incorporating an earlier proposal into the design.

\subsection{\textit{An Intrusion Response Approach for Elastic Applications Based on Reinforcement Learning~\cite{id23}}}

The paper~\cite{id23} serves as a continuation of~\cite{id20} previously discussed in this survey. Building upon the prior research, the current study expands and refines the approach by proposing an Intrusion Response Controller (IRC) based on deep RL and transfer learning. The focus remains on addressing the common limitation in existing approaches, which often neglect the non-stationary behavior of the protected system.

In this continuation, the researchers present a comprehensive evaluation and tuning of hyperparameters for NN, which form the core of the IRC. The integration of transfer learning proves to be a significant enhancement, reducing the transient adaptation stage when changes occur in the number of replicas of specific services.

A critical aspect that this paper emphasizes is the consideration of elasticity, a key characteristic of cloud computing platforms. The study demonstrates how the IRC effectively responds to changes in the number of instances of system components, leveraging the scalability offered by modern cloud computing. To assess the proposed approach, the researchers employ Online Boutique (OB) 2.0~\cite{onlineBoutique} as a reference system, a web application used by Google to showcase cloud-enabling technologies. The evaluation takes place at both the system and application levels, focusing on vulnerability assessment and agent training for self-protection.

 \rmvtxt{Experimental results further validate the efficacy of transfer learning, showcasing its ability to reduce the training time required for model adaptation when instances are removed. Moreover, the study demonstrates the speed-up achieved by the NN-based IRC, particularly in large-scale systems. This controller produces near-optimal sequences of defense actions by effectively utilizing the stateful model of the system based on the MDP framework.}

The contributions of this paper lie in its innovative approach to addressing the challenges posed by non-stationary environments and enhancing security in Software-Defined Networking (SDNs). By integrating deep RL and transfer learning, the proposed IRC offers an automated and efficient defense mechanism for cloud-based systems. This advancement eliminates the need for complete re-planning when system parameters change, ensuring timely and effective responses to emerging cyber threats.

The state structure, actions, and reward function in this paper follow a similar pattern to the previous work. The state structure of the system is defined by four variables: ``active", indicating the status of microservices (whether they are running or not); ``restarted", denoting whether a microservice has been restarted during the execution of a defense policy; ``compromised", marking if an attacker has gained control over a microservice; and ``shellCorrupted", signifying if an attacker has overwritten the shell/bin/sh in the container.

The actions set of the IRC consists of eight actions, each having the capability to modify the state of the microservice on which they are executed. These actions are characterized by their execution time, monetary cost, and pre-and post-conditions functions. The reward function in this approach is based on a penalty score, which takes into account the execution time and monetary cost associated with the selected action.

Presented in this paper is a methodology based on ML, serving the purpose of minimizing risk within elastic applications. An advanced RL evolution algorithm is utilized for the selection of appropriate countermeasures. The research highlights the potential of cutting-edge technologies to bolster network security and adaptability in dynamic network environments. The improvements presented in this continuation work compared to the previously analyzed paper underscore the dedication of researchers to advancing the field of IRS. In this work, a state-based system is again used for the representation of the system structure. This is a very interesting idea for the description of device states, which can help AI algorithms. Another advantage is the effectiveness of the use of transfer learning. This reduces the training time of the model, which in practice can result in models that can be upgraded faster as new threats appear.

\subsection{\textit{GTM: Game Theoretic Methodology for optimal cybersecurity defending strategies and investments~\cite{id7}}} 
The paper~\cite{id7} presents Game Theory Methodology (GTM), a comprehensive methodology that combines AG analysis with game-theoretic techniques~\cite{gametheory} to automatically predict possible attacking scenarios and propose optimal defending strategies. The primary objective of GTM is to achieve optimal allocation of the cybersecurity budget for effective cybersecurity risk mitigation by implementing effective security controls.

Specifically, AGs are utilized to model various attack paths, encompassing potential exploits that can be exploited by attackers. The methodology incorporates game theory principles to capture the behaviors and strategies of both attackers and defenders during cybersecurity incidents. In this context, the interactions between the attacker and defender are treated as a zero-sum game~\cite{zerosum}. Zero-sum game is a situation in which one person’s gain is equivalent to another’s loss, so the net change in wealth or benefit is zero. In addition, the application of the Nash equilibrium algorithm~\cite{nashequi} is employed to solve this game and derive optimal strategies. Through this approach, optimal defending strategies are determined to minimize the organization's cybersecurity risk.

GTM functions as a tool that will be divided into 3 modules: the Graphical Attack Engine, the Data Pool, and the Defence Strategy. 

The first module (i.e., the Graphical attack engine) uses MulVAL to generate AGs that model all possible attacking scenarios and paths of an organization. It receives as input a vulnerability assessment report that is the output of a vulnerability assessment tool (e.g., Nessus~\cite{nessus}). \rmvtxt{The produced graph consists of nodes representing logical propositions, where the source of an attacker's potential privileges is expressed as a propositional expression based on network configuration parameters. }

The Data Pool module serves as a centralized repository of guidelines, laws, and reports about cybersecurity and privacy. Its primary objective is to assist organizations in their defense against cyber-attacks. The module functions as both a database and a platform for ongoing enrichment by Chief Information Security Officers (CISOs). \rmvtxt{These CISOs actively contribute new input, ensuring that the Data Pool remains up-to-date and relevant to the evolving cybersecurity landscape.}

The last module, the Defence Strategy, is the most important, as it is responsible for calculating the expected loss and choosing the most appropriate cybersecurity controls based on a defined budget.  
The organization's loss (L) is calculated using Equation~\ref{equ:equation1}, which incorporates the probability of a threat occurrence to a specific asset (Pa) multiplied by the probability of successful exploitation (Psa), and further multiplied by the impact of that successful exploitation (I), derived from a business impact analysis
\begin{equation}\label{equ:equation1}
L = PA \times PSA \times I
\end{equation}

To emphasize the defensive aspect, the authors introduce a parameter representing the level of security (S). In Equation~\ref{equ:equation2} , the authors incorporate the summation of the losses associated with each involved asset of the organization. This addition expands the equation to encompass the cumulative losses across all assets, providing a comprehensive measure of the organization's overall losses.

\begin{equation}\label{equ:equation2}
L_o = \sum_{a\in A} L_a
\end{equation}
 
This paper outlines a methodology, grounded in Game Theory with AGs. Although the paper is not specifically focused on countermeasure selection, nash equilibrium is used as an algorithm for minimizing risk within the restrictions of a maximum cost. Therefore, the disadvantage of this document is the lack of the search for the selection of countermeasures. On a positive note, the proposal presented is very attractive. Focusing the use of the proposal towards countermeasure selection could be interesting. By making the games between attacker and defender more dynamic, and potentially incorporating 0-day vulnerabilities, the methodology could achieve even greater efficiency in decision-making by autonomous systems. Such an approach could be mixed with earlier ideas such as the dual mode for countermeasure selection: static and dynamic. Since GTM operation could be considered a slow process, and in a dynamic mode of countermeasure selection, speed is rewarded. Given these considerations, the lack of use of AI is also a slight negative aspect of the proposal. Indeed, the use of AI could benefit the proposal. For instance, in dynamic mode, the AI could identify vulnerabilities more effectively, while in static mode it learns from the data and prepares a stronger strategy. Consequently, this type of model could be used in a more static environment where all possibilities can be explored promptly without time constraints. 

\subsection{\textit{Security Countermeasures Selection Using the Meta Attack Language and Probabilistic Attack Graphs~\cite{id8}}} 

The paper~\cite{id8} introduces a comprehensive approach to countermeasure selection for enhancing the security of critical infrastructures. By utilizing the Meta Attack Language (MAL)~\cite{MAL} framework, the authors propose a methodology that allows for formalizing and addressing countermeasure selection problems effectively. \rmvtxt{The proposed algorithm, based on a mathematical model, offers flexibility and adaptability to different scenarios, enabling efficient decision-making processes for improving security.}

\rmvtxt{With MAL, IT infrastructure models can be created and reused, providing a structured representation of devices and their relationships. These models serve as the foundation for generating AGs, where nodes depict attack and defense steps, and edges represent the logical flow of attacks.}

The mathematical model employed for countermeasure selection takes an infrastructure model as input and generates a set of recommended defensive measures considering various constraints. These constraints include resource limitations, such as budget or time restrictions, dependencies between actions that must be executed in a specific order, and the ability to identify mutually exclusive measures that cannot be simultaneously implemented, such as cutting essential services.

The input to the model is an AG modeled with the MAL framework, and the output is a partially ordered set of all defense steps (D), where D comes from: v $\in$ D $\rightarrow$ Pre(v) $\subseteq$ D and w $\leq$ v for every w $\in$ Pre(v), where v and w are defense steps, and the relation w $\leq$ v means that w needs to be implemented before v. And Pre() function is a prerequisites function that describes temporal dependencies between defense steps. Also, this partially ordered set of defense steps must be within the limits of budget and time that are specified for each action and are expressed as a vector of the sum of all the costs in time and budget.

An iterative approach is employed to simulate attacks and select appropriate countermeasures. Each iteration involves simulating an attack while assuming that the previously generated solution's defense steps are implemented. Critical steps in the attack paths are identified, and corresponding countermeasures are examined. When a defense step is found that satisfies all constraints, it and its prerequisites are added to the solution. This process continues iteratively, updating the set of countable steps as new defenses are incorporated. \rmvtxt{By iteratively refining the solution, the research aims to effectively enhance the security of the system while addressing all necessary constraints.}

In summary, the paper introduces an AG-based methodology approach. For the purpose of minimizing risk in critical infrastructure, the authors employ a mathematical algorithm for countermeasure selection. The use of the MAL framework as an approach for infrastructure modeling can be seen as an advantage due to the accurate representation that this type of framework can provide.  Two main disadvantages are identified in this research. On the one hand, the proposal is designed almost exclusively for a static mode of defence. On the other hand, the authors do not use AI in the proposed framework. The integration of AI techniques could improve the adaptability of the solution to dynamic scenarios and allow the detection of unknown attacks. Combining the modelling capabilities of MAL with the mathematical model and AI for a dynamic approach could provide a more stable solution. Two noteworthy advantages are identified. The first relates specifically on the ability to consider the possibility of adding constraints, such as budget or time constraints. The other advantage has to do with the consistency obtained from the persistence of the defenses of the previously generated solution. In other words, it is very interesting to maintain consistency in the selection of countermeasures by storing the previously selected solutions.

\subsection{\textit{Optimal Defense Strategy Selection Algorithm Based on Reinforcement Learning and Opposition Based Learning~\cite{id10}}} 

The paper~\cite{id10} presents a hybrid strategy differential evolution algorithm that combines RL and opposition-based learning. This algorithm is proposed to construct the optimal security strategy for ICS. By integrating these advanced techniques, the paper addresses common problems associated with evolutionary algorithms and aims to significantly enhance the effectiveness and efficiency of the optimization process.

The authors propose a framework for optimal security strategy selection consisting of three steps: risk assessment, construction of a target attack function benefit-cost of protection, and selection of the optimal security strategy. The risk assessment starts by modeling the cyber-attack using an AG, which is then transformed into a Bayesian AG by calculating the time probability of successful exploitation of each vulnerability. \change{To weigh the benefits and costs, the attack benefits and protection costs are quantified, and an objective function is established. The objective function aims to minimize the attack benefits and protection costs while ensuring that the protection cost remains within the allowable limit.}{Benefits and costs are quantified to establish an objective function aiming to minimize attack benefits and protection costs within allowable limits.}

For the selection of the optimal security strategy, the differential evolution algorithm is employed as a population-based optimization technique. \change{It evolves a population of Np D-dimensional parameter vectors through mutation, crossover, and environmental selection operations. The algorithm starts by generating initial test solutions randomly, aiming to find the optimal security strategy by improving the population of solutions based on evolutionary principles.}{It iteratively evolves a population of Np D-dimensional parameter vectors through mutation, crossover, and selection operations, aiming to improve solutions based on evolutionary principles.} The selection operation determines which test solutions enter the next-generation population as candidate solutions. The iterative process of mutation, crossover, and selection continues until all experimental solutions are selected or until the accuracy of the current solution meets the requirements or a certain number of iterations is reached.

A hybrid differential evolution algorithm, called RODE, is proposed to leverage this strategy.  In RODE, RL is employed, where the agent learns through a process of ``trial and error" by interacting with the environment and receiving rewards or punishments based on its actions. Q-learning, a popular RL method, is used to determine the best actions based on the received rewards. Additionally, opposition-based learning is incorporated to enhance the convergence speed and escape local optima. This is particularly important as the population may reach a point where it becomes trapped in a locally optimal solution.

This paper presents a methodology based on AGs to minimize cost and risk in ICS. For this purpose, an evolution algorithm, that operates with RL, is used for the selection of the optimal security strategy. In this paper, the combination of different AI techniques within a hybrid framework proves to be highly effective in analyzing and finding optimal solutions. Particularly, the inclusion of Opposition-Based Learning is a clever addition that allows the framework to explore alternative paths and avoid getting trapped in local optima. However, it is important to note that this solution primarily focuses on building security policies and preventive security strategies. It excels in solving static problems and providing effective measures against known attacks. Yet, its limitations lie in addressing dynamic attacks and the utilization of dynamic countermeasures. One could say that exploring ways to enhance the framework's adaptability to handle evolving and sophisticated attack scenarios, thus further improving its effectiveness in real-time threat mitigation would be beneficial.

\subsection{\textit{An Intrusion Detection System for IoT Using KNN and Decision-Tree Based Classification~\cite{id11}}} 

The paper~\cite{id11} proposes the application of DL techniques to address the limitations of traditional IDS and ML algorithms in handling large volumes of data~\cite{deep11.1}. The focus of the study revolves around threat identification within the IoT context. To analyze the data flow and protect IoT devices, the authors employ two ML classifiers: K-Nearest Neighbors (KNN) and Decision Tree (DT). These classifiers are utilized to assess various metrics such as error rate, accuracy, precision, recall, and F1 score.

The proposed methodology begins with the initial step of selecting a dataset, followed by the conversion of categorical values into numerical data through preprocessing techniques. In this particular study, a dataset comprising 25,000 rows and 25 columns was utilized for both training and testing purposes. The subsequent stage involves performing normalization operations on the data. The dataset is then split into two subsets, with 67\% of the rows allocated for training and the remaining 33\% designated as the test set. These datasets are subsequently employed to evaluate the performance of the KNN and DT classification models. Various performance metrics such as Error Rate, Accuracy, Precision, Recall, and F1 are utilized to assess the effectiveness of the models.

The model yielded near perfect results in terms of the five evaluation metrics used, indicating its high effectiveness in detecting threats in the DoH20 dataset~\cite{DoH20}.  The results demonstrate promising outcomes, and the data preprocessing for classification is executed accurately. 

This paper discusses a methodology founded on ML, designed for detection of threats within an IoT environment. Employing DL with supervised learning as an algorithm for completing the task. For starters, the positive aspect of this paper has to do with the ranking models used: KNN and DT, as well as the number of metrics that have been used to understand the performance of the proposal. And, although the focus of this paper does not specifically address the selection of countermeasures using AI, such systems must possess the capability to identify potential threats within the system. Exploring this aspect of threat identification can prove valuable in enhancing overall security measures. However, for this proposal, there is a limitation regarding the availability of information on the model's reproducibility and adaptability to different environments with varying conditions and parameters. In addition to this, there is a disadvantage based on the inability to explain these AI models. This can present a challenge when it comes to decision making by an agent. Where it is not possible to explain why the output of the autonomous system is the way it is. 

\subsection{\textit{Discovering Exfiltration Paths Using Reinforcement Learning with Attack Graphs~\cite{id13}}} 

In~\cite{id13} the use of AGs, RL, and cyber terrain to identify optimal paths for data exfiltration in enterprise networks is explored. Building upon previous research on crown jewels identification, which aimed to compute optimal paths for compromising critical assets, this work takes an inverted approach by assuming that data has already been stolen and focuses on quietly exfiltrating it from the network. RL is employed to develop a reward function that identifies paths with reduced detection to support the exfiltration process.

In RL agents learn by taking actions in environments and receiving rewards, these environments are modeled using MDP from AGs. MDP are tuples [$S$, $A$, $\phi$, $P$, $R$] where S and A are states and actions, $\phi$ are admissible state-action pairs, P is the probability transition function and R is the expected reward function, and the agent interacts with an environment $\epsilon$ = [$S$, $A$, $\phi$, $P$, $R$] by taking actions and receiving states and rewards. RL serves as a targeted tool for cyber operators to improve the efﬁciency of operator workﬂow in penetration testing.

The paper introduces a service-based defensive terrain model in CVSS-MDP, leveraging CVSS scores to estimate the probability of successful exploitation. This approach allows the model to emulate human-like attack behaviors. The services are categorized into authentication, data, security, and common, and a hierarchical cost structure is established for attacking these services. For example, authentication has a negative reward of -6, data has a reward of -4, and both security and common have a reward of -2 for exploiting actions. Conversely, scanning actions receive a reward increase of 1.

The method employs multiple terminal states representing different exit nodes of interest and a single initial node. The agent interacts with the network in an episodic manner to determine the optimal exit node based on expected rewards. \rmvtxt{The tool provides a comprehensive path analysis for cyber operators to identify the best exit node.}

For the purpose of threat detection within a system, this paper introduces a methodology relying on AGs that employs an algorithm based on RL. The main disadvantage of this approach lies in the lack of focus on reaction and countermeasure selection. Additionally, the use of AGs can suffer from scalability issues in networks with many devices~\cite{attackgraphlimitation}. Such a limitation of the use of AGs has to be taken into consideration before using this approach. However, the proposal can be interesting for the reason that it can lead to the intelligence process that a reaction system could additionally have. This feature would be to identify when there has been an exfiltration of information, or if an information leak is taking place. On the positive side, it is worth commenting that by modeling the service-based defensive terrain and incorporating CVSS scores, the approach allows for a more accurate estimation of the probability of successful exploits. This information can serve as a valuable resource for reaction systems and to anticipate and counteract the next steps of cybercriminals. Another worth commenting advantages is the ability of the tool to provide a comprehensive path to operators. This element can be seen as a way of explaining how the AI algorithm has solved the problem. The combination of explainability, coupled with the application of the RL algorithm concept to the realm of reactions, could serve as a compelling factor in designing an effective system.

\subsection{\textit{Cyber-attacks detection in industrial systems using artificial intelligence-driven methods~\cite{id14}}} 

In~\cite{id14}, an innovative approach for detecting and identifying cyber-attacks in SCADA systems is introduced. The proposed method utilizes a stacked DL technique, along with eleven ML models, to enhance the accuracy and sensitivity of IDS. Real datasets from laboratory-scale SCADA systems are employed to evaluate the performance of the investigated methods, considering various evaluation metrics. 

The eleven ML models are Xtreme Gradient Boosting (XGBoost), RF, Bagging, support vector machines with different kernels, classification tree pruned by the minimum cross-validation and by 1-standard error rule, linear discriminate analysis, conditional inference tree, and the C5.0 tree.

Five Deep Neural Networks (DNN) were trained using the provided training data. The stacked DL model combines the predictions of these five trained networks through majority voting. Each of the five networks consists of three layers, with varying numbers of hidden neurons: (80, 60, 60), (80, 80, 60), (100, 80, 80), (120, 100, 80), and (180, 120, 80). These three-layer networks demonstrate the ability to accurately discriminate malicious attacks without overfitting the training data. The rectified linear units function is employed as the activation function for the hidden layers, while the softmax function is used for the output layers, addressing the multiple-class cyber-attack detection problem.

In the study, five metrics of effectiveness were employed to quantify the performance of all the methods, where each metric provides valuable insights into different aspects of the model's performance: Accuracy, Precision, Recall, F1-Score, and Area UnderCurve (AUC). 

Accuracy is a measure of overall correctness, indicating how well the model classified instances correctly. Precision focuses on the accuracy of positive predictions, evaluating the model's ability to avoid false positive predictions. Recall, also known as sensitivity, measures the model's capability to correctly identify positive instances and avoid false negative predictions. The F1-Score combines precision and recall into a single metric, providing a balanced measure of the model's performance. The AUC is a metric specific to binary classification models. It evaluates the model's ability to rank instances correctly, by plotting the true positive rate against the false positive rate at various classification thresholds.

The analysis of the research highlights the superiority of the XGBoost approach compared to other methods, demonstrating its exceptional detection performance. The advantageous characteristics of XGBoost, such as its ability to avoid overfitting, handle outliers robustly, and maintain invariance to feature transformations, contribute to its outstanding results. It is worth noting that the DL models, despite their capability to capture complex feature interactions, did not achieve the best performance in this specific case study. 

To conclude, a methodology based on ML models has been used in this paper. Furthermore, to detect threats in critical infrastructures, DL has been used as an algorithm between other techniques such as Random forest (RF), XGBoost, and Bagging for comparison. As in previous analyses, the presented research does not contain the selection of countermeasures desired, but it does propose a study of the different ML models. It is these results that can be very relevant when choosing models for reactive systems, which can be considered an advantage. The reason for this consideration is that it can help future researchers to quickly select ML models that are suitable for their specific case. However, it is important to consider the limitations of supervised classifiers, including the requirement for prior knowledge of different attack types to achieve efficient detection. Constructing these ML models relies on labeled training data, making it challenging to encompass all existing attack types, especially with the emergence of advanced and sophisticated attacks on a daily basis. Finally, a further identified disadvantage could be mentioned. The drawback concerns the explainability part of the ML models. In this paper it is not specified that any of the models present this kind of feature, however in this particular proposal, without a framework or tool that translates the reasoning behind each model, it does not seem possible to explain each given solution.

\subsection{\textit{An Intrusion Response System utilizing Deep Q-Networks and System Partitions~\cite{id18}}}

In~\cite{id18} an open-source software prototype called ``irs-partition", which implements an IRS is presented. The software leverages Deep Q-Networks (DQN) and RL techniques to address the challenges posed by the non-stationary behavior of computer systems. By employing transfer learning, it can adapt and evolve with the changing nature of the system.

One of the main features of the IRS-partition prototype is its multi-agent formulation, which overcomes the curse of dimensionality by partitioning the protected system. This partitioning allows different local modeling techniques and solvers to be used, such as those based on MDP like DQN and Dynamic Programming. Each partition of the defended system is assigned an Independent Response (IR) agent responsible for its control.

The defended system is segmented into multiple subsystems, with an IR agent assigned to oversee each of them. Each agent plays a crucial role in achieving the overall system goal of maintaining security. They work towards predicting the near-optimal action for their respective partitions using a customizable DQN. These IR agents receive attack information from an IDS, which is responsible for analyzing data collected by sensors in the system partitions. It is assumed that the IDS component is present.

The IRS generates its response solely based on the system model, without using an attack model. This approach allows the system to effectively handle zero-day attacks, even though the response may be less precise compared to an IRS that relies on an attacker model for known attacks.

\rmvtxt{A system contains components of different types. These component types form partitions, representing sets of components belonging to the same type.}The system operates on states, actions, and rewards, where the reward function considers execution time and cost.

The software execution commences by creating the system model, which is then decomposed into partitions. For each partition, a DNN is generated, and a corresponding agent is trained. These agents are responsible for providing the local near-optimal next action based on the current partition state. \rmvtxt{The set of predicted optimal local actions collectively leads to a global optimum.}

To train the agents, the software utilizes DQN with Monte Carlo simulation~\cite{montecarlo}. The simulation starts with an initial system state set by the system administrator. Actions are executed in the environment based on this initial state, resulting in rewards and the subsequent system state. The agent selects actions by either exploiting the acquired knowledge to maximize the expected discounted reward or exploring actions with unknown outcomes in terms of reward and transition. The epoch continues until it terminates, either when the environment reaches a secure state or another defined stopping condition is met.
 
In conclusion, the paper presents a ML-based methodology. With the purpose of mitigating an attack, an evolution algorithm such as RL is used. Several readings can be drawn from this proposal. To begin with, the paper shows how the use of DQN and RL techniques can address the challenges posed by the non-stationary behavior of computer systems. In addition, it is a very interesting idea to use several agents to achieve the same goal, that is, to maintain security. Another positive point is that the IR agents that this prototype works with are trained to use information received directly from an IDS. This means that it can be used in a more realistic environment. As long as it is taken into account that the IDSs trained use the same data modeling as those used within an organization. Another advantage that has been identified is the use of Monte Carlo simulation, a normal forecasting model that gives very good results. The incorporation of these ideas can contribute to the development of preventive security management strategies.

\subsection{\textit{Cost-damage analysis of attack trees~\cite{id15}}} 

\cite{id15} focuses on the optimization of damage within a specified cost budget using Attack Trees (AT)~\cite{schneier1999attack}. AT are widely employed in the field of cybersecurity to model potential attacks on systems. Attackers aim to maximize the inflicted damage while adhering to a predefined budget constraint. The interplay between two crucial attack metrics, attack cost (representing the attacker's budget) and attack damage (indicating the harm caused to the system), is carefully considered.

AT provides a valuable framework for analyzing and strategizing potential attacks, enabling a systematic approach to identify vulnerabilities and mitigate risks. While damage is the primary concern for system owners, understanding the associated cost of an attack provides valuable insights into the feasibility and likelihood of such attacks.

\rmvtxt{The paper addresses the optimization problem by proposing solutions for both deterministic and probabilistic scenarios, considering different levels of uncertainty. Techniques such as ILP and bottom-up approaches are employed to develop effective methodologies for determining the maximum achievable damage within the given budget.}

Additionally, the paper presents three problem statements: finding the most damaging attack given a cost budget, finding the cheapest attack given a damage threshold, and finding the cost-damage Pareto Front. One notable challenge in cost-damage analysis lies in the fact that existing methods often overlook attacks that do not activate the top node of the attack tree, despite their potential to cause significant damage to intermediate nodes. \rmvtxt{The paper highlights the importance of considering these intermediate attacks in the analysis.}

The contributions of the paper can be summarized as follows: a formal definition of cost-damage problems in AT, proof of the NP-completeness of these problems, evidence that cost-damage problems cannot be reduced to common extensions of the binary knapsack problem, a bottom-up method to solve deterministic and probabilistic cost-damage problems for tree-structured AT and an ILP-based method to solve deterministic cost-damage problems for directed acyclic graph-like AT.

The paper introduces two novel methods for addressing cost-damage problems in AT, optimizing damage or cost under a constraint, and calculating the cost-damage Pareto Front. For tree-structured AT, a more efficient bottom-up approach is employed, while general AT benefit from the application of ILP. These methods can be extended to analyze other related problems in the cost-damage tradeoff.

In conclusion, this paper uses AT methodology and efficient algorithms such as heuristic probabilistic methods for solving cost-damage problems. Although the paper does not specifically focus on countermeasure selection systems, it introduces an intriguing concept with practical implications. The findings can be used for understanding security metrics and decision-making processes in the context of budget-constrained attacks. These methods and AT may be deployed in a reaction system for countermeasure selection based on budget. In addition, another advantage of this research has to do with the generation of different attack possibilities. This presents an interesting point of view as it generates information that can be very valuable for the planning of defensive strategies. This can be especially helpful for organizations that have limited budget. By doing so, these organizations could be able to prioritize their resources more effectively and protect the most probable attack scenarios. However, it is important to underline the limitations of AT. Such as the scalability limitations of ADTrees, AT present the same problems as they are based on trees~\cite{treeBased}. 

\section{Comparative Exploration of Analyzed Works}\label{ComparativeExploration}

In this section of the paper, a comparison of the research conducted is carried out. The main objective of this review is to provide a detailed clear overview of the current state of the art in threat response systems that use AI. To achieve such a goal, the comparative Table~\ref{table:analyzedFeatures} is used, which contains the analyzed contributions and highlights each of the features in Section~\ref{Features}. This table effectively visualizes the differences and similarities in techniques between the selected papers, providing a deeper understanding of the current state of research in this field.

Based on Table~\ref{table:analyzedFeatures} several significant trends and patterns can be identified in the analyzed articles. Therefore, the following subsections contain a review of each of the features and a brief analysis of their values. 

\begin{landscape}

\begin{table}[]
\centering
\footnotesize
\begin{tabular}{|l|l|p{3cm}|p{5cm}|p{4.5cm}|p{3cm}|l|} \hline 

{\textbf{ID}} & {\textbf{Year}} & {\textbf{Methodology}} & {\textbf{Algorithm}}& {\textbf{Strategy Objectives}}& {\textbf{Measurement and results}}& {\textbf{Use Case}}\\ \hline  
 \cite{id4}                                & 2018                               & Fault tree Analysis& Fuzzy Theory                                                             & Vulnerability Assessment& -                                                                 &Multiple                                    \\ \hline  
 \cite{id5}                                & 2018                               & Attack Graphs& Mathematical Model& Minimizing Risk& -                                                                 &Private Network\\ \hline 
\cite{id1}                               & 2019                               & Attack Graphs redefined& A* algorithm& Minimizing Risk& Computing time                                                    & Multiple                                      \\ \hline  
\cite{id2}                                & 2019                               & Machine Learning& Neural Network - Self-taught Learning& Threat Detection                                    & Variety of Classification& Multiple                                    \\ \hline  
 \cite{id24}                                & 2019                               & Bayesian Network& Pareto Optimal Solutions& Minimizing Risk& Countermeasure Selection Speed&Critical Infrastructure\\ \hline  
\cite{id3}                                & 2020                               & Hypergraphs                               & StRORI                                                                   & Minimizing Risk and Cost& Countermeasure Selection Speed/Estimated Risk/Losses& Multiple\\ \hline  
\cite{id9}                                & 2020                               & Attack Graphs& Attack Defense Trees& Minimizing Risk& Countermeasure Selection Speed& Multiple                                    \\ \hline  
\cite{id17}                                & 2020                               & Bayesian Network& Probabilistic Model& Minimizing Risk& Implementation Cost and Impact& Multiple                                    \\ \hline  
 \cite{id19}                                & 2020                               & Dynamic Countermeasure Knowledge& -                                                                        & Countermeasure Standard Proposal& ROI Score                                                         &Multiple                                    \\ \hline  
 \cite{id20}                                & 2020                               & Machine Learning& Evolution Algorithm - Reinforcement Learning (Markov   Decision Process)& Minimizing Risk& Positive Rate&Microservices Architecture\\ \hline  
 \cite{id21}                                & 2020                               & Hierarchical Risk Correlation Tree        & Autonomous Response Controller                                           & Minimizing Risk& Countermeasure Selection Speed&Cyber–Physical Systems                   \\ \hline  
 \cite{id16}                                & 2021                               & Machine Learning& Artificial Immune Systems                                                & Minimizing Risk& Countermeasure Selection Speed&Multiple                                    \\ \hline  
 \cite{id23}                                & 2021                               & Machine Learning& Evolution Algorithm - Reinforcement Learning& Minimizing Risk& Positive Rate&Elastic Applications                           \\ \hline  
 \cite{id7}                                & 2022                               & Game Theory& Nash Equilibrium& Minimizing Risk and Cost& -                                                                 &Private Network\\ \hline  
 \cite{id8}                                & 2022                               & Attack Graphs& Mathematical Model& Minimizing Risk& Countermeasure Selection Speed&Critical Infrastructure\\ \hline  
 \cite{id10}                                & 2022                               & Attack Graphs& Evolution Algorithm - Reinforcement Learning& Minimizing Risk and Cost& Positive Rate&Industrial Control Systems\\ \hline  
 \cite{id11}                                & 2022                               & Machine Learning& Deep Learning - Supervised Learning                                      & Threat Detection& Positive Rate&Internet of Things                                       \\ \hline  
 \cite{id13}                                & 2022                               & Attack Graphs& Evolution Algorithm - Reinforcement Learning& Threat Detection& Positive Rate&Multiple                                    \\ \hline 
\cite{id14}                                & 2022                               & Machine Learning& Deep Learning& Threat Detection& Positive rate                                 & Critical Infrastructure\\ \hline  
\cite{id18}                                & 2022                               & Machine learning                          & Evolution Algorithm - Reinforcement Learning& Mitigate Attack& Positive Rate& Multiple                                    \\ \hline  
\cite{id15}                                & 2023                               & Attack Trees& Heuristic - Probabilistic methods                                        & Calculation of Cost-damage probabilities in Attacks& Calculation Speed& Multiple                                    \\ \hline 
\end{tabular}%
\caption{Side-by-side comparison of the selected works based on the features described in Section~\ref{Features}}
\label{table:analyzedFeatures}
\end{table}

\end{landscape}

\subsection{Methodology}

As can be seen, the two most frequent values for the methodology observed in Table~\ref{table:analyzedFeatures} are ``Machine Learning" and ``Attack Graph". These two values can show a trend followed by researchers in this field. Clearly in Section~\ref{Section:Survey}, these two methodologies produce very good results which is why they are widely adopted. On the one hand, ML offers a wide range of algorithms that have been developed. These algorithms provide great flexibility when it comes to solving different problems. In addition, they are algorithms that are very popular and receive a lot of interest from the community~\cite{upriseML,upriseML2}. This supports their usefulness and effectiveness in many contexts. Specifically, in the context of reactive systems, ML presents some relevant advantages. To begin with, such a methodology can manage a very large amount of data efficiently, which means it has a strong scalability component. Furthermore, in an autonomous reactive system, it can be used to automate countermeasures after the detection of attacks, as well as learn from the received data and evolve to face changing threats. In other words, it also presents an interesting component of adaptability. 

On the other hand, AGs are used in detailing standard attack patterns and calculating risks across systems. Furthermore, graphs are a very powerful tool that has been shown to have different advantages. The understandability of AGs is of significant importance, particularly considering their highly visual nature. If an autonomous system were to base a decision on these graphs, it is imperative that operators can easily understand their meaning and implications. Additionally, they are an interesting tool when planning incident response plans, in which risk probabilities can be used to help focus resources on specific assets.  

However, AGs present limitations~\cite{aglimitation}, as well as problems of scalability and time to generate them. Due to globalization and growth in the number of devices, scalability, and performance issues could greatly affect proposals using these methodologies. On the ML side, there are different drawbacks~\cite{MLlimitations}. For example, ML requires a strong database for both training and evolution. This may present challenges in organizations with little data, or data of poor quality. Another disadvantage has to do with the lack of interpretability in the decision-making process of this type of methodology. This can make operators hesitant to carry out the proposed countermeasures. 

\addtxt{From a practical point of view, one can say that it makes sense to employ machine learning-based methodologies, especially in enterprise environments. In these enterprise networks, where numerous vulnerabilities may exist, a large amount of data is constantly being generated. Each of these points can provide a vast amount of information that needs to be processed and analyzed to provide effective countermeasures against threats detected. ML therefore can extract meaningful patterns and make accurate predictions from these large datasets. It could proactively analyze threat alerts generated by IDSs and seek appropriate countermeasures. Furthermore, ML has the advantage of being able to adapt and continuously improve over time as more information is provided.}

\addtxt{In the application of graph-based methodologies, within organizational settings such as corporate or military environments, several significant benefits for risk management and security can be identified. These methodologies can provide a clear visual representation of potential attack routes that could compromise security. This allows security teams to quickly understand potential threats and infrastructure hotspots. Similarly, it allows for the representation of multiple attack scenarios, providing a more complete understanding of potential threats and their possible impacts on the organization.}

\subsection{Algorithm}

From the column corresponding to the ``Algorithm" feature, one can easily notice that the values are not as concentrated as they were in the methodology column. This implies that researchers are exploring and experimenting with a variety of algorithms for problem-solving. That is due to the fact that academics are attempting to find the optimal approach to the problem of implementing AI-based reaction systems. 

This diversification in the amount of algorithms may negatively affect the development of an autonomous reaction system. That is because it may prevent researchers from following a common strategy. Furthermore, this variety could present various problems when integrating components into organizations or make it harder to compare the various proposals due to their differing similarities.

\addtxt{From a pragmatic standpoint, the diversity of algorithms found in the papers analyzed can be attributed to several reasons. On the one hand, different problems and research contexts require specific algorithmic approaches to solve them. For example, while data classification problems may benefit from DL algorithms such as NN, optimization problems may require evolutionary or game theory-based algorithms. The variety of algorithms may reflect the diversity of tools available in the fields of data science and AI. Researchers can select the most appropriate algorithm according to their prior knowledge, experience, and the nature of the problem they are tackling. This diversity of tools may result in the choice of different algorithmic approaches to similar problems, depending on individual preferences and computational resource constraints.}

\addtxt{This diversity reflects the multifaceted and constantly changing nature of these fields, where innovation and experimentation are fundamental to the advancement of knowledge and technology, and where the lack of research in the field of reactive systems has not laid a foundation for the creation of robust reactive systems that work in an enterprise environment.}

The importance of NNs is also highlighted, derived through the case studies analysis of~\cite{id2} and~\cite{id23}. It has been proven that NNs have a significant impact on threat identification and response in various environments. However, the importance of proper training is emphasized, as poor training can lead to crucial omissions in the detection and response process.

\subsection{Strategy Objectives}

In the case of the ``Strategy Objectives" column, it is readily observable that the most frequent objective is risk minimization. This is achieved through the selection or generation of countermeasures to counteract threats. \change{The risk minimization objective can also include values in which the costs associated with implementing countermeasures are taken into account.}{However, it is important to note that risk minimization can include values in which the costs associated with implementing countermeasures are taken into account. } These reflects the goals and needs perceived by researchers in the field, where the objective is to optimize cybersecurity investment without compromising the protection of critical systems and assets. \addtxt{Nevertheless , it is worth questioning whether this pursuit of cost minimization could lead to compromises in the effectiveness of countermeasures and ultimately in system security.}

\addtxt{Furthermore , the usefulness of AI in this case lies in its adaptability to the strategy chosen for each case. The application of AI methods, such as machine learning models, may vary depending on the specific threat and the organization's security objectives. Ultimately, integrating A into cyber security strategies could not only improve response capabilities, but also facilitate a more flexible response to evolving cyber threats depending on the strategy the organization chooses to pursue.}

Additionally, there is an effort to strike a balance between mitigating identified threats and the resources available at that moment. \addtxt{This search for efficiency in resource allocation is essential to ensure an effective response to cyber threats, but it may also pose significant challenges. For example, it could generate tensions between the need to implement rapid countermeasures and the limited availability of financial or technical resources.}

A noteworthy aspect of the analysis and the table is~\cite{id16,id19} where the objective is to establish a standard for countermeasures. Building on this premise, and based on the survey, it can be seen that there is no universal standardization process attempting to address these solutions uniformly. Instead, there are several specific proposals for each model, particularly concerning countermeasures. This diversity of approaches has resulted in a fragmented landscape of solutions that, each with its unique characteristics, do not demonstrate a trend toward convergence. Although this diversity could present an intriguing opportunity to discover unconventional and previously unexplored solutions, it is important to note that most works do not seek to unify the various research towards a common objective.

Concerning countermeasures, in many articles, it is observed that they are ``predefined". In other words, countermeasures that are inflexible and lack the capacity to adapt to the context. This can represent a problem when it comes to dynamizing the solution of the reactive system in almost real-time. Additionally, it can prevent the identification of unknown threats, where a more defined system might not find a response.

\subsection{Measurements and results}

In relation to the column ``Measurements and Results", it can be seen that there is a limited range of values associated with the performance tests. The data demonstrates predominantly the trend toward the metric ``Positive rate”.  This trend is particularly significant in the context of the primary challenge at hand, which revolves around the selection of appropriate countermeasures to mitigate potential threats effectively. This metric serves as a valuable indicator of how well the chosen countermeasures align with the specific threats they are intended to address.

\change{Additionally, it is worth noting the use of metrics such as computation time or countermeasure selection time. These metrics shed light on the efficiency and speed at which the countermeasures are identified and implemented. This efficiency is particularly relevant in environments that face threats, where rapid response times can make a significant difference in containing potential damages and minimizing risks.}{Moreover, the emphasis on metrics such as computation time and countermeasure selection time underscores the importance placed on efficiency and speed in the context of response systems. While rapid response times are indeed crucial for mitigating potential threats, an overemphasis on speed may inadvertently overshadow other critical factors, such as the accuracy and efficacy of the selected countermeasures. This drawback raises concerns about whether the evaluation criteria adequately capture the multifaceted nature of cybersecurity challenges and response strategies.}

\addtxt{In this context, AI integration could offer competitive advantages. AI systems could help in streamlining automation tasks by taking into account the different ways of measuring the results that a system could have. Trying to adjust different system configurations to get better results depending on the metrics used.  }

\change{To conclude the analysis of this column, it may be interesting to note that during the survey, many risk measures could be identified in terms of vulnerabilities, such as CVSS or CVE, while for countermeasures no value has been found that could classify them.}{The absence of comprehensive risk measures for countermeasures is a notable gap in the analysis. While vulnerabilities can be assessed using established measures such as CVSS or CWSS, a corresponding framework for assessing countermeasures seems to be missing. This shortcoming undermines the ability to effectively compare and prioritize countermeasures based on their potential impact and their effectiveness in mitigating specific threats. It also highlights a more general problem in the domain, where the focus on identifying vulnerabilities or exposures often exceeds the focus on developing and evaluating robust countermeasures.}

\addtxt{In conclusion, while the analysis provides valuable insights into certain aspects of performance and efficiency, it also reveals underlying limitations and areas for improvement. A more balanced and comprehensive approach to performance assessment, covering a wider range of parameters and taking into account the effectiveness of countermeasures to mitigate various threats, is essential to move forward.}

\subsection{Use Case}

When examining the ``Use Case" column, it becomes evident that the value ``All" holds significant prominence. This indicates that most of the works analyzed in this survey do not tailor their solutions to a specific system. \change{Consequently, approaches marked with this value possess the advantage of adaptable deployment across diverse environments, ensuring highly efficient protection for most scenarios.}{While this approach offers the advantage of versatile deployment in a variety of contexts, allowing for effective protection in a range of scenarios, it also raises critical questions about the suitability of these one-size-fits-all solutions.}
\change{However, it is important to consider that by not tailoring the proposal to a specific system, certain specific areas within different environments may remain unprotected. For instance, in critical infrastructure environments, meticulous consideration of all details is imperative, given the heightened significance of these devices compared to common devices.}{By opting for a generic approach, there is a risk of overlooking the complex and nuanced requirements and vulnerabilities inherent in different systems and environments. This lack of customization may leave certain areas within various environments inadequately protected, especially in high-risk sectors such as critical infrastructure. In these environments, where the consequences of a cyber attack can be catastrophic, a more tailored and meticulous approach to cyber security may be required.}

\addtxt{Furthermore, the prevalence of the term ``All'' may indicate a tendency to oversimplify or generalize in the development and evaluation of cyber security solutions. While adaptive deployment is undoubtedly valuable, it must be accompanied by a thorough understanding of the specific risks and challenges posed by different use case scenarios. This could lead to a false sense of security and leave critical systems vulnerable.}

\addtxt{In conclusion, while the preference for the designation ``All'' in the ``Use case'' column highlights the adaptability of the methodologies or systems analyzed, it also underlines the need for a more critical and tailored approach to the development of cybersecurity solutions.}

\section{Research challenges}\label{Section:ResearchChallenge}

Based on the findings described in the previous sections, potential future research directions for further progress in the field of AI-driven threat response systems can be analyzed. The next subsections explore this potential future research direction to set out a road map as a possible future path for researchers.

\subsection{Unified Frameworks}\label{UnifiedFrameworks}

One of the primary research challenges lies in establishing standardized procedures and unified structures within AI-powered threat response systems. The current challenge with these systems is caused by the lack of a common baseline. This lack of shared foundations can have several significant implications and consequences. \rmvtxt{For instance, absence of interoperability between tools or systems that have to be coordinated for the exchange of information. Alternatively, it may make it difficult to compare different proposals because they have no common basis for bench marking.}

\addtxt{The absence of a shared baseline results in a significant lack of interoperability between tools and systems, preventing a potential fluent exchange of information between researchers and security teams in practical scenarios. . This fragmentation poses significant challenges when comparing different proposals, as there is no standardized benchmark for assessment.}

\change{By developing unified frameworks that provide a shared basis for the integration of various AI techniques, consistency and interoperability could be introduced into these response systems. Such an approach could accelerate development processes and pave the way for collaborative advances that could strengthen cohesion among researchers and practitioners.}{To overcome this obstacle, it is imperative to develop unified frameworks. Such frameworks would outline standardized protocols and structures for the seamless integration of various AI techniques, thereby promoting consistency and interoperability. The implementation of such frameworks would streamline the development process and foster collaboration between researchers and practitioners.}

The challenge of establishing unified frameworks implies the creation of standardized protocols and structures that integrate a variety of AI techniques seamlessly. Researchers could work towards developing a comprehensive taxonomy of AI methodologies and threat response approaches that would provide a common language and understanding. 

\subsection{Collaborative AI}\label{collaborativeAI}

Collaboration between different AI methodologies, especially NN and ML~\cite{id2,id22}, stands out as a promising field for further exploration.  The concept of combining multiple AI techniques, can be explored for potentially relevant breakthroughs. By bringing together these diverse techniques, integrated systems can be developed to leverage the strengths of each approach. This integration purpose is to take advantage of the strengths of each one, thereby achieving the highest precision and capacity to counter threats that arise.

To improve collaboration between different AI methodologies, researchers could focus on creating a standardized platform that facilitates the test of the integration of different AI techniques. \change{In this case, different methodologies could be assessed and ranked according to their advantages and weaknesses based on specific tasks. }{Through this platform, different methodologies could be evaluated and ranked according to their effectiveness for specific tasks, taking into account their respective strengths and weaknesses. }Subsequently, \change{complex}{sophisticated } systems could be designed that unify models according to their strength, thus, \change{building}{creating }  hybrid models that design the reaction problems and assign tasks to different AIs. 

It is worth remarking that this research challenge is closely related to the previous one, related to the unification of frameworks. Exploring the concept of a common base structure that not only shares a common set of inputs but also facilitates the creation of hybrid AI models could significantly simplify both model development and testing processes. This approach would simplify and create potential synergies. It could bring together researchers and practitioners in the search of improvement of model creation and testing in order to strengthen AI-powered threat response systems. 

\subsection{Evolutive AI}\label{AI Evolution}

Adaptation and evolution mechanisms of AI models used in threat response systems represent another interesting research perspective. Due to the dynamic nature of cyber threats, AI models require evolution and adaptation. This is because threats are evolving, and new and unknown vulnerabilities are being discovered regularly. \change{Then, similarly, as threats evolve, AI models must be dynamically updated. }{This requires real-time adaptation of AI models to remain effective over time and to avoid possible obsolescence.}This challenges presents an intriguing research trajectory to facilitate real-time updates driven by the evolving threat landscape. The real-time evolution would allow AI models to remain effective over time and could avoid the obsolescence that some of the models in this survey may exhibit.

\change{It is important to note that while AI can be designed to be evolutive, it can also present a series of potential issues. Among these is the need to re-train models, which could present problems of time and loss of prior knowledge. Moreover, this re-training would have to be programmed, and this presents another dilemma in terms of determining the optimal time for this evolution. }{Implementing evolutive AI presents practical challenges, particularly concerning re-training models. Re-training can be time-intensive and may result in the loss of prior knowledge, impacting operational efficiency. Additionally, determining the optimal timing for model updates is essential to ensure that systems are responsive to emerging threats without disrupting ongoing operations. To address these challenges practically, automated mechanisms for model updates are crucial. Establishing a robust feedback loop between real-world threat data and AI models enables regular and timely updates.}

\rmvtxt{To address this challenge, automated mechanisms should be designed to update AI models. The implementation of a feedback loop between real-world threat data and AI models can ensure regular and timely updates.}

\rmvtxt{Alternatively, techniques such as transfer learning and RL~[104], [119], [135] can be applied to facilitate faster adaptation of AI models to new threat scenarios.}

\addtxt{Furthermore, practical techniques such as transfer learning and} RL~\cite{id10,id12,id20} can \addtxt{expedite the adaptation process. Transfer learning allows models to leverage knowledge from pre-trained models, reducing the need for extensive re-training from scratch. RL, on the other hand, can simulate various threat scenarios and enable AI models to learn and adapt in real-time based on feedback from the environment. }

\change{Besides, cooperation with threat intelligence providers can be established to provide such threat data. This would allow access to updated information for adapting the models. }{Collaboration with threat intelligence providers is another practical strategy. Partnering with these providers can provide direct access to the latest threat information, enabling faster and more informed decision-making for model adaptation. By integrating real-time threat feeds into AI systems, researchers can ensure that models are updated with the most current threat data. }


\subsection{Collaborative Knowledge Exchange}\label{CollaborativeKnowledge}

To achieve effective cybersecurity, collaboration is a strong foundation. Collaboration allows for an approach that promotes research towards collaborative learning. By enabling organizations and states to collaborate in sharing information/knowledge for training AI models without compromising data privacy, a collective defense mechanism could be developed and, due to the joint effort, become potentially more effective. Particularly, this line of research could trigger the generation of shared knowledge leading to a more robust and informed defense against evolving threats.

To address this challenge, the \change{first}{initial} step should be to establish trusted frameworks and protocols for secure collaboration\change{. To share threat data securely without revealing sensitive information.}{, ensuring the sharing of threat data without compromising sensitive information.} \change{Next}{Subsequently }, efforts should be made to find an effective way of sharing information that reaches widely and equally. For example, a centralized platform that could act as a threat intelligence intermediary. 

This concern also raises certain data-related issues. In the case of detection models, the datasets that can be found are abundant~\cite{dataSetDetection1,dataSetDetection2}, since it is a more investigated area, and the data for training is also easier to obtain due to its nature. Nevertheless, when it comes to training the different reaction models, there is hardly any data to train and test the performance. Which presents a problem, since AI models base their knowledge on the quality and quantity of the datasets on which they are trained. This could also be considered within this challenge, as it has to do with the data and the way it is exchanged. 

This challenge is connected to several previous challenges. First, it is connected to the sub-section~\ref{UnifiedFrameworks} since a common framework would require collaboration to support information sharing. As a second connection, the sub-section~\ref{collaborativeAI}  is also related, in the same way as the previous one, in order to achieve collaborative AI, data needs to be shared. 

\subsection{AI and Human Experts}\label{AIandHumanExperts} 

In the context of collaboration is not limited to the organizational environment. Indeed, effective collaboration between humans and AI systems presents another intriguing scenario. Future research could focus on designing intuitive user interfaces that facilitate interaction between human experts and AI-based threat response systems. This collaboration empowers decision-makers by providing them with AI-generated knowledge while allowing for human oversight and judgment, which improves the overall threat response process. \rmvtxt{In addition, this also addresses the need to explain the decisions made by autonomous response systems, named as eXplainable Artificial Intelligence (XAI)[138], requiring the technicians operating these systems to have control over the decisions being made.}

\addtxt{A practical approach to confront this challenge involves designing user interfaces that prioritize usability and user experience. Researchers can conduct usability studies and apply human-centered design principles to create intuitive interfaces that foster cooperation between humans and AI. An essential aspect of this collaboration is the integration of eXplainable Artificial Intelligence (XAI) techniques}~\cite{XAI}. 
 \addtxt{XAI enables AI systems to explain their decisions in a transparent and understandable manner, allowing human experts to validate and interpret AI-generated insights.  }

To confront this challenge, researchers could conduct usability studies and apply human-centered design principles to create user-friendly interfaces that enhance cooperation between AI and humans. In addition, incorporating explainable AI techniques can build trust by enabling human experts to understand and validate AI-generated decisions. An example of such a scenario would be the AI4CYBER~\cite{AI4CYBER} project, which is working on AI-based security solutions in order to facilitate the work of operators and developers. Furthermore, this project is focused on AI detection and testing as well as in the field of reaction. 

\subsection{Countermeasure standardization}

As a final alternative to the proposed research challenges, the topic of predefined countermeasures emerges. As mentioned in the survey discussion, many articles present ``predefined" countermeasures for response systems~\cite{id3,id5,id8,id23}. That is, countermeasures that are not dynamically generated by a reactive system, but have a static and unchanging base which has to be used for all scenarios.  As a result, a challenge arises concerning the dynamic nature of searching for a more universally applicable solution. It may therefore be attractive to extend this line of research in pursuit of more dynamic and adaptable approaches to maintain a countermeasure database. The use of artificial intelligence may be interesting to generate new countermeasures based on the identification of emerging vulnerabilities or threats.

To effectively standardize countermeasures, efforts should be dedicated to the categorization of threats and vulnerabilities in a standardized way. Developing a common language for countermeasures that fits this categorization can help standardize responses~\cite{id16,id19}. \addtxt{By establishing a common language for countermeasures that fits this categorization, responses can be standardized across organizations and industries. This not only facilitates communication and collaboration, but also ensures consistency and interoperability in the application of security measures.} 

\change{In addition, leveraging ML and natural language processing to analyze and classify emerging threats can automate the generation of countermeasures. }{In addition, the integration of advanced technologies such as ML and natural language processing  can significantly improve the effectiveness of countermeasure standardization efforts. ML algorithms can analyses large amounts of threat data to identify patterns and trends, enabling automatic classification of emerging threats. Similarly, NLP techniques can extract valuable information from unstructured data sources such as security reports and threat intelligence sources, further enriching the threat analysis process. By leveraging these technologies, organizations can streamline the generation and deployment of countermeasures. }

\rmvtxt{To effectively standardize countermeasures, little efforts should be dedicated to the categorization of threats and vulnerabilities in a standardized way. Reduced efforts can be attributed to the fact that, as mentioned above, there are standards for the representation of vulnerabilities and threats that are widely used, such as CVE and CVSS.  This implies that progress should continue to be made in this field, however}

\addtxt{However, while there are existing standards such as CVE and CVSS for vulnerability representation, progress should continue in this field. }It could be \change{interesting}{beneficial} for academia to research a method for standardization of countermeasures that provides a scoring. In the same way as the categorization of vulnerabilities, such a score could be used to assess the effectiveness of a countermeasure. Thus helping the optimal selection of countermeasures by a reaction system. In essence, developing a common language for countermeasures that fits this categorization can also help standardize responses~\cite{id16,id19}. 

However, this also requires collaboration, as industry stakeholders need to agree on the standard to be followed. Not only that, but they must apply it to new developments and try to update old systems. 

\subsection{Generative AI}


In 2017, the article ``Attention is All You Need"~\cite{AttentionIsAllYouNeed} presented the Transformer architecture for artificial intelligence. A model that allowed the network model to learn the relationships between parts of an input stream. In 2018, ``Improving Language Understanding with Unsupervised Learning"~\cite{Radford2018ImprovingLU} was presented, an article that presented BERT, an improvement that allowed bidirectional models to be trained by capturing contextual information in both directions in a sequence of text, allowing context to be understood. In 2019, the OpenAI team launched GPT-2~\cite{ReleaseStrategies}, increasing the size of the model and extending the performance of tasks such as natural language processing. In 2020, GPT-3~\cite{GPT3} was released, further increasing its performance and creating ChatGPT. After extensive study, and to the best of our knowledge, the use of this type of artificial intelligence, called generative modeling, has not been put into practice for the reactive systems environment. Several research papers have been found that use this type of model for malware detection~\cite{DetectionGenerative,DetectionGenerative2}. However, no papers have been found that attempt to combine generative models for reactive systems.  

Such models could be used in a variety of ways. For a start, it could be used to create an interface that simplifies the work of a cybersecurity operator. This advantage is related to the ``AI and Human Experts" challenge, as it could extend the explainability of AIs, as well as simplify and accelerate the tasks of incident response or threat prevention. An additional potential benefit could be the automatic creation or reading of incident reports, facilitating the process of information sharing. This would be linked to the ``Collaborative Knowledge Exchange", where generative models could be on both sides sharing information and breaking it down as it is received. As a final example, a generative model could be used for threat simulation through text generation, so that the security configurations implemented in a system could be checked and tested automatically. 

\addtxt{A significant aspect of this challenge lies in the integration of generative AI models into existing cyber security frameworks. Unlike traditional reactive systems, which rely on predefined rules and signatures to detect and mitigate threats, generative AI models have the ability to autonomously generate new attack scenarios and vulnerabilities. This dynamic nature introduces complexities in threat detection and response.}

\addtxt{
Furthermore, the ethical implications of using generative AI models in cybersecurity cannot be overlooked. The use of AI-generated adversarial examples, for example, raises concerns about the potential for unintended damage or exploitation. Investigating the ethical considerations and implications of integrating generative AI models into reactive systems is therefore crucial to ensure responsible and equitable cybersecurity practices.}

In conclusion, the link between generative artificial intelligence models and reactive systems in the world of cybersecurity presents a great challenge that could be very interesting to investigate.  

\subsection{Summary}

\begin{figure}[h!]
    \centering
    \includegraphics[width=\columnwidth]{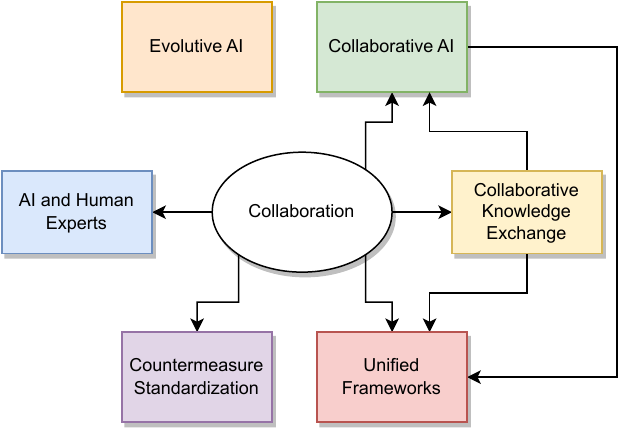}
    \caption{\change{Graph with research challenge interconnections.}{Graph showing the interconnections of the research challenges according to their similarity and their dependence on collaboration.}}
    \label{fig:researchSummary}
\end{figure}

To enhance the reader's understanding, an illustrative support, represented by \figurename~\ref{fig:researchSummary}, has been incorporated. This figure visually presents the research challenges, illustrating their interconnections through arrows. In particular, an additional parameter called ``Collaboration" has been introduced to indicate whether a research pathway requires stakeholder collaboration. \addtxt{The introduction of ``Collaboration" underscores the importance of cooperative efforts in addressing these complex and multifaceted challenges.}

\addtxt{For instance, developing ``Collaborative AI" requires AI systems to work together and share information. This inherently requires collaboration among different AI developers, researchers, and possibly end-users to ensure that these systems can interact and learn from each other efficiently. Similarly, ``Collaborative Knowledge Exchange" involves the continuous flow of information and insights between AI systems and human experts, requiring a collaborative framework to facilitate this exchange.}

\addtxt{A central element in this network of research challenges is the ``Unified Frameworks" challenge. This challenge is pivotal as it connects with ``Collaborative Knowledge Exchange" and ``Collaborative AI," emphasizing the need for comprehensive frameworks that can integrate different AI systems and facilitate seamless interactions. }

\addtxt{``Countermeasure Standardization" challenge also demands collaboration as it involves creating standardized approaches to mitigate risks associated with AI systems.``AI and Human Experts" focuses on the integration and synergy between AI systems and human expertise. Lastly, ``Evolutive AI" stands as an independent challenge that does not directly require collaboration with other research pathways. However, its development can still benefit from collaborative insights, particularly in understanding how AI systems can evolve and adapt over time.}

\section{Conclusions and future work}\label{Section:Conclusion}

Human dependence on the use of information technologies has grown exponentially. This dependence poses considerable security challenges, which can be seen in frequent cyber-attacks targeting businesses, critical infrastructures, and companies, and even impacting geopolitical events. In response, both private enterprises and nation-states are actively exploring effective ways to strengthen the security measures of their systems, aiming to safeguard sensitive data and prevent unauthorized access. This trend has triggered an increase in research efforts to address these critical challenges. The emergence of cutting-edge technologies, like AI, has enriched the landscape. When integrated into the cybersecurity sector, this technology offers numerous advantages, such as the identification of threats and the identification of appropriate countermeasures. With the assistance of AI, dedicated systems can be used to detect, prevent, and respond to threats within the realm of cyberspace.

This paper presented a novel survey that delves into the recent advancements within the domain of AI-driven threat response systems. The primary objective is to closely examine and compare diverse recent papers in this area, trying to highlight and uncover the various complexities associated with integrating AI into cyber defense. Through this exploration, the survey identifies opportunities to strengthen defense mechanisms using AI-powered response systems, paving the way for future research.

The survey presented focuses on examining several research studies that share common characteristics. From this analysis, some interesting approaches that use a variety of AI techniques to solve the challenge of creating an autonomous threat response system are presented. The variety of approaches found reflects the range of the field and the growing diversity of strategies for dealing with cyber threats. 

\addtxt{However, the survey has also identified certain gaps in the research. One of the most notable is the diversity of methodologies and algorithms presented in each paper. This lack of standardization poses a major challenge, as it makes it difficult to compare the different papers analyzed.}

\addtxt{From a practical point of view, this lack of uniformity can complicate interoperability. In scenarios such as the creation of a federated CERT (Computer Emergency Response Team), where collaboration and information sharing are essential, the diversity of methodologies and algorithms could obstruct the integration and sharing of data between the different entities. To address this challenge, more attention is needed on standardization and research collaboration. Adopting common standards and encouraging collaboration between researchers can help establish best practices and facilitate comparison and knowledge sharing in the field of cyber threat reaction.}

It is important to note that this comparison reflects a snapshot in time and that the field of threat response systems using AI continues to develop. Future research advances and discoveries could have an impact on this comparative survey, further improving our understanding of this essential area of research.

Concerning the methodological target, one of the key aspects is the selection of countermeasures to minimize both the risk and the costs associated with the identified threats. The research highlights the importance of this strategic choice and how AI techniques can influence the determination of the most appropriate countermeasures. There is also an emerging trend in the integration of machine learning techniques and Neural Networks to improve the accuracy of threat identification. 

While not the primary focus of this literature review, it has also incorporated studies presenting solutions that can be adapted or evaluated for response systems. These additional works provide a complementary perspective on how to address the various challenges and can serve as a basis for future research in this evolving area. As cyber threats continue to evolve, collaboration between different approaches and disciplines becomes critical to developing robust and effective threat response strategies.

Finally, a research challenge section has been proposed in which different future research directions have been suggested. Among these, work will be done investigating the development of some interesting concepts such as: 

\begin{itemize}
    \item[--]  The use of generative models for the creation of useful interfaces for cybersecurity operators.
    \item[--]  Designing automated mechanisms to update AI models in response to real-world threat data.
    \item[--] Create a standardized platform for testing the integration of different AI techniques.
    \item[--] Create a comprehensive taxonomy of AI methodologies and threat response approaches.
\end{itemize}

\section*{Acknowledgment}
This work has been partially funded by the strategic project CDL-TALENTUM from the Spanish National Institute of Cybersecurity (INCIBE) and by the Recovery, Transformation and Resilience Plan, Next Generation EU, and from the postdoctoral grant Margarita Salas (172/MSJD/22).

\bibliographystyle{IEEEtran}
\bibliography{ref.bib}{}

\end{document}